\documentclass[aps,prb,twocolumn,amsmath,showpacs,superscriptaddress]{revtex4}

\usepackage{graphicx}

\usepackage{color}
\usepackage{bm}
\usepackage{amsmath}
\usepackage{amssymb}

\newcommand{\VTG}{V_{\mathrm{TG}}}
\newcommand{\VPG}{V_{\mathrm{PG}}}
\newcommand{\mum}{\mu\mathrm{m}}

\begin{document}
\title{Effects of Coulomb screening and disorder on artificial
graphene based on nanopatterned semiconductor}

\author{O. A. Tkachenko}
\affiliation{Rzhanov Institute of Semiconductor Physics of SB RAS, Novosibirsk, 630090 }

\author{V. A. Tkachenko}
\affiliation{Rzhanov Institute of Semiconductor Physics of SB RAS, Novosibirsk, 630090 }
\affiliation{Novosibirsk State University, Novosibirsk, 630090, Russia}

\author{I. S. Terekhov}
\affiliation{School of Physics, University of New South Wales, Sydney 2052, Australia}

\author{O. P. Sushkov}
\email{sushkov@phys.unsw.edu.au}
\affiliation{School of Physics, University of New South Wales, Sydney 2052, Australia}


\begin{abstract}
A residual disorder in the gate system is the main
problem on the way to create artificial graphene based on two-dimensional
electron gas. The disorder can be significantly screened/reduced due to
 the many-body
effects. To analyse the screening/disorder problem we consider
 AlGaAs/GaAs/AlGaAs heterostructure with two metallic gates.
We demonstrate that the design least susceptible to the disorder
corresponds to the weak coupling regime (opposite to tight binding)
which is realised via system of quantum anti-dots.
The most relevant type of disorder is the area disorder which is
a random variation of areas of quantum anti-dots.
The area disorder results in formation of puddles.
Other types of disorder, the position disorder and the shape disorder,
are practically irrelevant.
The formation/importance  of puddles dramatically  depends
on parameters of the nanopatterned heterostructure.
A variation of the parameters by 20--30\% can change the relative amplitude of
puddles by orders of magnitude.
Based on this analysis we formulate criteria for
the acceptable design of the heterostructure aimed at
creation of the artificial graphene.
\end{abstract}


\maketitle

\section{Introduction}
Graphene  is a carbon monolayer with a hexagonal honeycomb lattice. The
material has  a very special band structure described by the
ultrarelativistic Dirac equation. Graphene has a  number of exceptional
properties that originate from  the quantum and ``relativistic'' nature of the
material.  Discovery of this material in 2004 has had a profound impact on condensed-matter physics \cite{Novoselov07,CastroNeto09}
and technology \cite{Novoselov12}.
Electrons in graphene can travel for long distances without scattering, the
electron mobility  is 10 times larger than that in Si at room temperature.
The major reason for the high mobility of electrons in graphene
is  the very high quality of the carbon lattice, there are very few defects
in the lattice.
An additional reason for the high mobility is the ``ultrarelativistic'' physics.
It is well known that there is no  backscattering for ultrarelativistic particles and this enhances the mobility.

The extraordinary properties of graphene originating from the particle dynamics
 in honeycomb/triangular lattice, have stimulated a number of researchers to
realize  'quantum simulators' of graphene  in other systems. We broadly refer to
this direction as artificial graphene. The major advantage of artificial
graphene is in larger degree of control. These systems are tunable and therefore can be driven
to complex topological/correlated phases which are not possible in
natural graphene. Study of these phases is of fundamental importance
and can lead to novel electronic applications such as the spin based
transistor.

Recent advances have demonstrated the possibility of creating artificial graphene
in diverse quantum systems. Current methods of design
and synthesis include (i)~nanopatterning of ultrahigh-mobility two-dimensional
(2D) electron
gases (2DEGs) \cite{Singha,park09,Gibertini09,simoni10,Nadvornik12,Goswami12},
(ii)~molecule-by-molecule assembly of honeycomb lattices on
metal surfaces by scanning probe methods \cite{Gomes12},
(iii)~trapping ultracold fermionic and bosonic atoms in honeycomb optical
lattices \cite{Wunsch08,Soltan,Tarruell12,Uehlinger},  and
(iv)~confining photons in honeycomb photonic crystals \cite{haldane,Bittner:2012,Rechtsman:2013}.
For a review of the recent advances see Ref.~\onlinecite{polini13}.

It can be argued that whereas the molecular, atomic and photonic analogues of
graphene offer a very high quality of the artificial lattice, the nanofabricated
semiconductor analogue enables the exploration of the impact of
long-range interactions, many-body effects, as well as the spin orbit
interaction \cite{Sushkov13}.
Technologically the semiconductor route to artificial graphene
offers the key advantage of scalability as silicon and III-V materials
are suitable for conventional top-down nanofabrication approaches.
The realization of artificial graphene in semiconductors requires a fine
control over disorder,  whereas the other analogues of graphene are either disorder
free (cold atoms on optical lattices) or characterized by tunneling
energy scales that are substantially larger than disorder (molecules
deposited on metals).

The residual disorder in heterostructure and in the gates system is the main
problem on the way to create 2DEG artificial graphene suggested in
Refs.~\onlinecite{Singha,park09,Gibertini09,simoni10,Nadvornik12,Goswami12}.
In the present work we analyze the role of disorder and Coulomb screening
in 2DEG artificial  graphene.
As a result of the analysis we formulate  criteria  for quality of quantum
engineering necessary
to create the artificial graphene.
Lattices of anti-dots and quantum dots are investigated
for more than 20 years, see
Refs.~\onlinecite{Weiss1991,Baskin92,Dorn04,Gibertini09,mrres2012,Nadvornik12}
 and references therein, but observation of a miniband structure
remains elusive.
The main reason for that is a relatively poor quality of the
created superlattice potential. The amplitude of uncontrolled random
deviations of the  potential from the perfect one are larger than the
characteristic scale of the miniband structure.
The honeycomb/triangular superlattice for artificial 2DEG
graphene can consist of an array of quantum
dots~\cite{Gibertini09} or, alternatively, of an
array of quantum  anti-dots \cite{Nadvornik12,TkaTka14}.
The present analysis indicates that  the anti-dot structure is less
sensitive to the superlattice
disorder since the minibands are more dispersive. Therefore, here we pursue the
``anti-dot'' route.
Usually systems of quantum dots and antidots are created in the 2DEG formed by doping.
In simpler systems, such as quantum point contact, an alternative approach
has been exercised, where the 2DEG and the nanosystem are patterned by two gates
without doping. In this case the mobility of the low-density 2DEG is
drastically increased
due to the suppressed impurity scattering \cite{Pfeiffer,Harrell,pyshkin}.
Therefore,  we consider an undoped
AlGaAs/GaAs/AlGaAs heterostructure with two metallic gates.
The lower metal gate laying on the surface of the semiconductor has
perforation with period 100--130\,nm, the gate is biased with a positive attracting
voltage. A voltage on the top gate creates the anti-dots in the 2DEG beneath
the holes perforated in the lower gate.

The paper is organized as follows. In Section \ref{Section1} we consider
an ideal device without any disorder in perforated gates. Even this problem
is pretty involved technically due to the Coulomb screening in the gates and
the self-screening in 2DEG. We find conditions for stability of the Dirac point,
values of voltages which have to be applied to the gates, the miniband
structure, the density of states, and the conductance of a finite
size ``sample''.
In Section \ref{Section2} we analyze disorder in the gates, in particular
random deviations of sizes/positions/shapes of quantum anti-dots from the
perfect structure. Here we calculate the density of states and the
conductance of a finite size ``sample''.
Hence we determine the critical degree of disorder to preserve the Dirac
physics.
Our results and conclusions are summarized in Section~\ref{sec:conclusions}.

\section{Ideal superlattice\label{Section1}}
\subsection{Laterally patterned heterostructure}
\begin{figure}[ht]
\includegraphics[width=0.2\textwidth,clip]{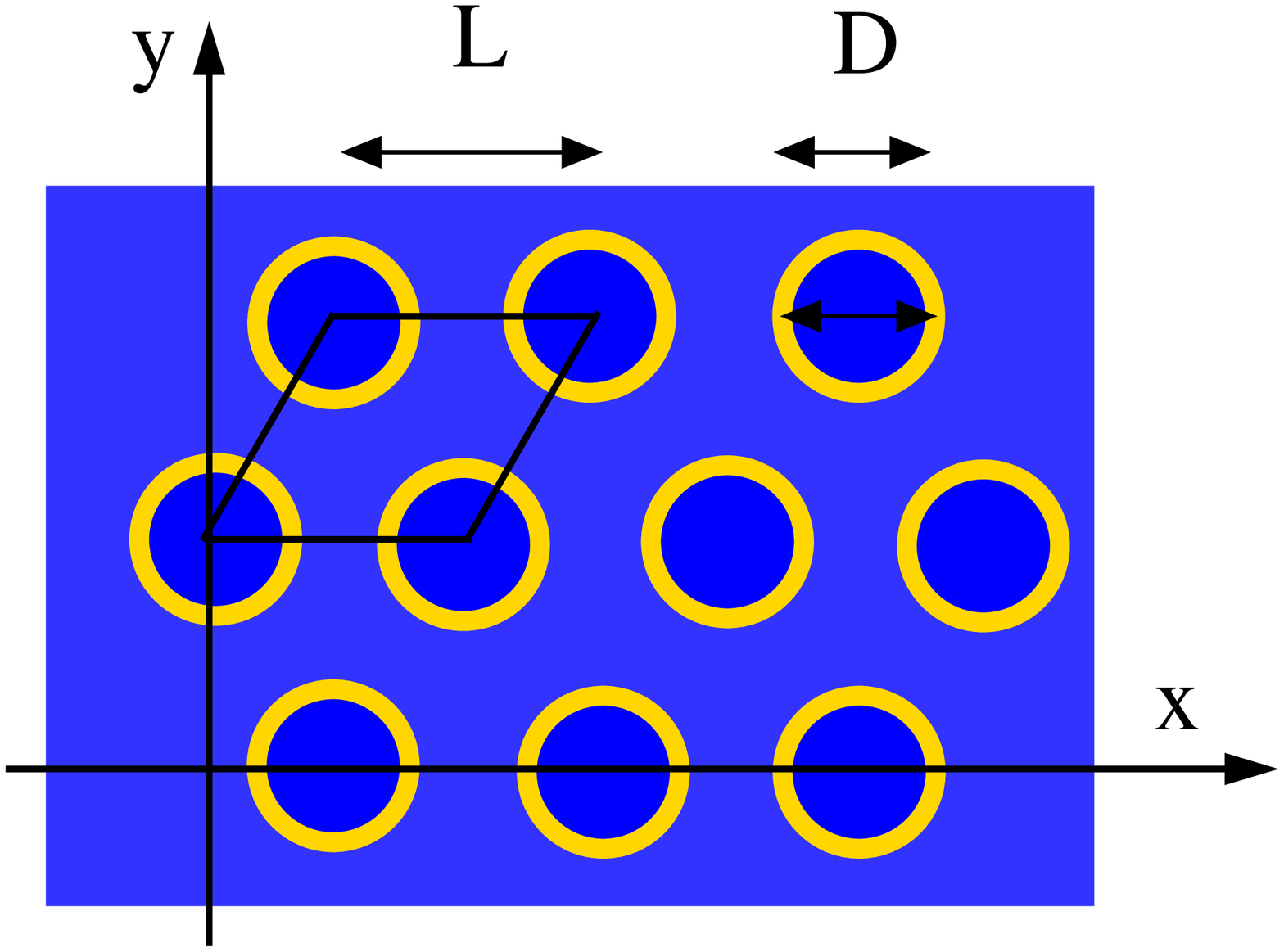}
\includegraphics[width=0.24\textwidth,clip]{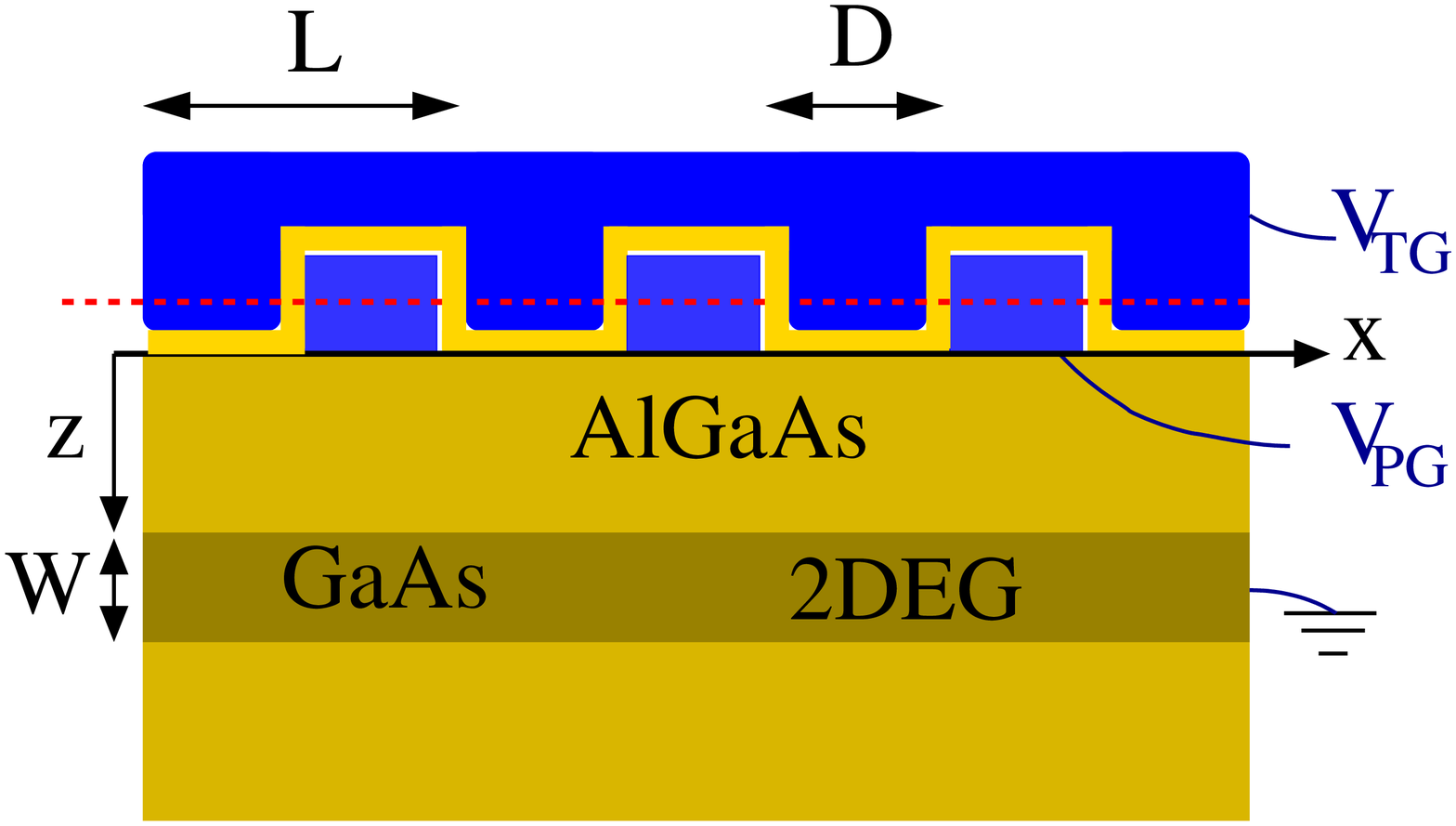}
\caption{Laterally patterned heterostructure.
The top and the perforated metallic gates are shown
by blue colors of slightly different intensity to distinguish the gates.
The gates are under voltages $\VTG$ and $V_{PG}$ respectively.
$L$ is the lattice spacing of the gate superlattice.
$D$ is the diameter of the perforation.
The left panel shows a horizontal cut indicated by the red dashed line
in the right panel.  Yellow and golden colors indicate an insulator,
AlGaAs and GaAs layers respectively. The 2DEG confined within the GaAs
layer is grounded.
$W$ is the width of the 2DEG quantum well.
}
\label{tri}
\end{figure}

\begin{figure}[hb]
\includegraphics[width=0.15\textwidth,clip]{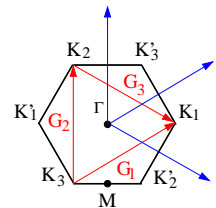}
\caption{Brillouin zone of the triangular superlattice.
}
\label{BZ}
\end{figure}

The laterally paterned heterostructure which we consider is
shown in Fig. \ref{tri}.
Brillouin zone of the triangular superlattice with lattice spacing  $L$
is shown in Fig.~\ref{BZ}. Wave vectors of the reciprocal lattice,
\begin{eqnarray}
\label{trvi}
{\bm G}_1=\frac{2\pi}{3L}(3,\sqrt{3}), \ \
{\bm G}_2=\frac{2\pi}{3L}(0,2\sqrt{3}), \ \
{\bm G}_3={\bm G}_1-{\bm G}_2,\nonumber
\end{eqnarray}
are also shown in Fig.~\ref{BZ}.
The points ${\bm K}_1$, ${\bm K}_2$, ${\bm K}_3$
are connected by vectors ${\bm G}_i$, and ${\bm K}_i'$
are obtained from the ${\bm K}_i$ by reflection.
Our convention is that $z=0$ corresponds to the bottom of the perforated gate,
see Fig.~\ref{tri}. The electrostatic potential created by the gates,
is a periodic function of $\bm\rho=(x,y)$ which satisfies the Laplace equation,
hence
\begin{eqnarray}
\label{u}
&&U_G(\bm \rho,z)=\mathrm{const}-eEz+\sum_{mn\ne 00}v_{mn}e^{-k_{mn}z}\cos(\bm k_{mn}\cdot\bm \rho)
\nonumber\\
&&\bm k_{mn}=m\bm G_1+n\bm G_2\,.
\end{eqnarray}
At $z=0$, close to the gates, the Fourier expansion  has all harmonics.
However, away from the gates, higher harmonics decay much faster than the
first harmonic and hence in the 2DEG plane the potential created by gates is
\begin{eqnarray}
\label{u1}
&&U_G(\bm \rho) \approx \mathrm{const}+
2U_0\sum_{i=1}^3\cos\left(\bm G_i\cdot\bm \rho\right)
\\
&&U_0\approx e \frac{D (\VTG-\VPG)}{2L}J_1\left(\frac{2\pi D}{\sqrt{3}L}\right)
\exp\left\{-\frac{4\pi z}{\sqrt{3}L}\right\}\ .\nonumber
\end{eqnarray}
Here $e$ is charge of the electron,
$z$ corresponds to the location of 2DEG, and $J_1(x)$ is the Bessel
function.
For the anti-dot array, which we consider, $U_0$ is positive.
Negative $U_0$ corresponds to the array of dots.

\subsection{Minibands in the noninteracting electron approximation}
\begin{figure}[ht]
\includegraphics[width=0.2\textwidth,clip]{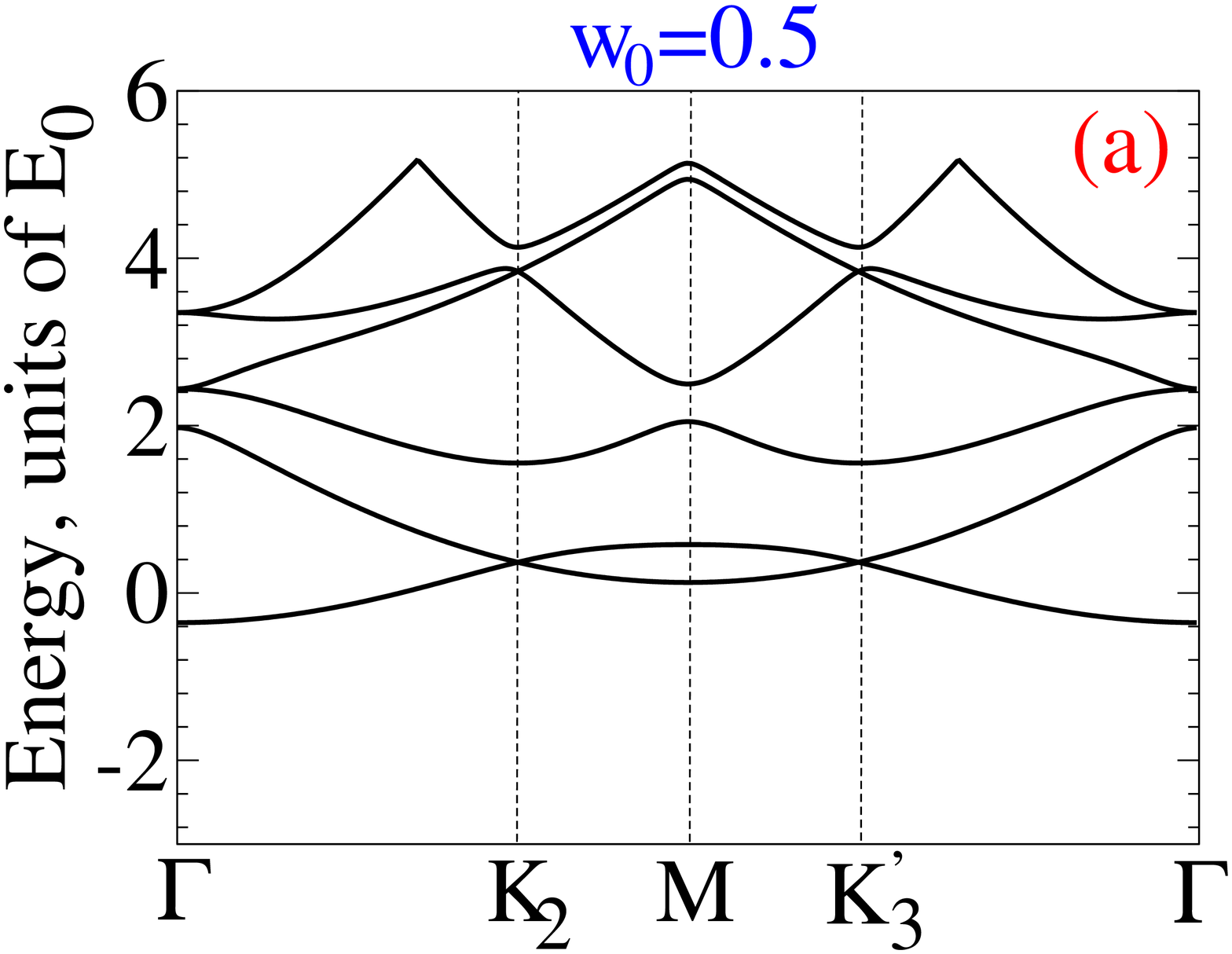}
\includegraphics[width=0.2\textwidth,clip]{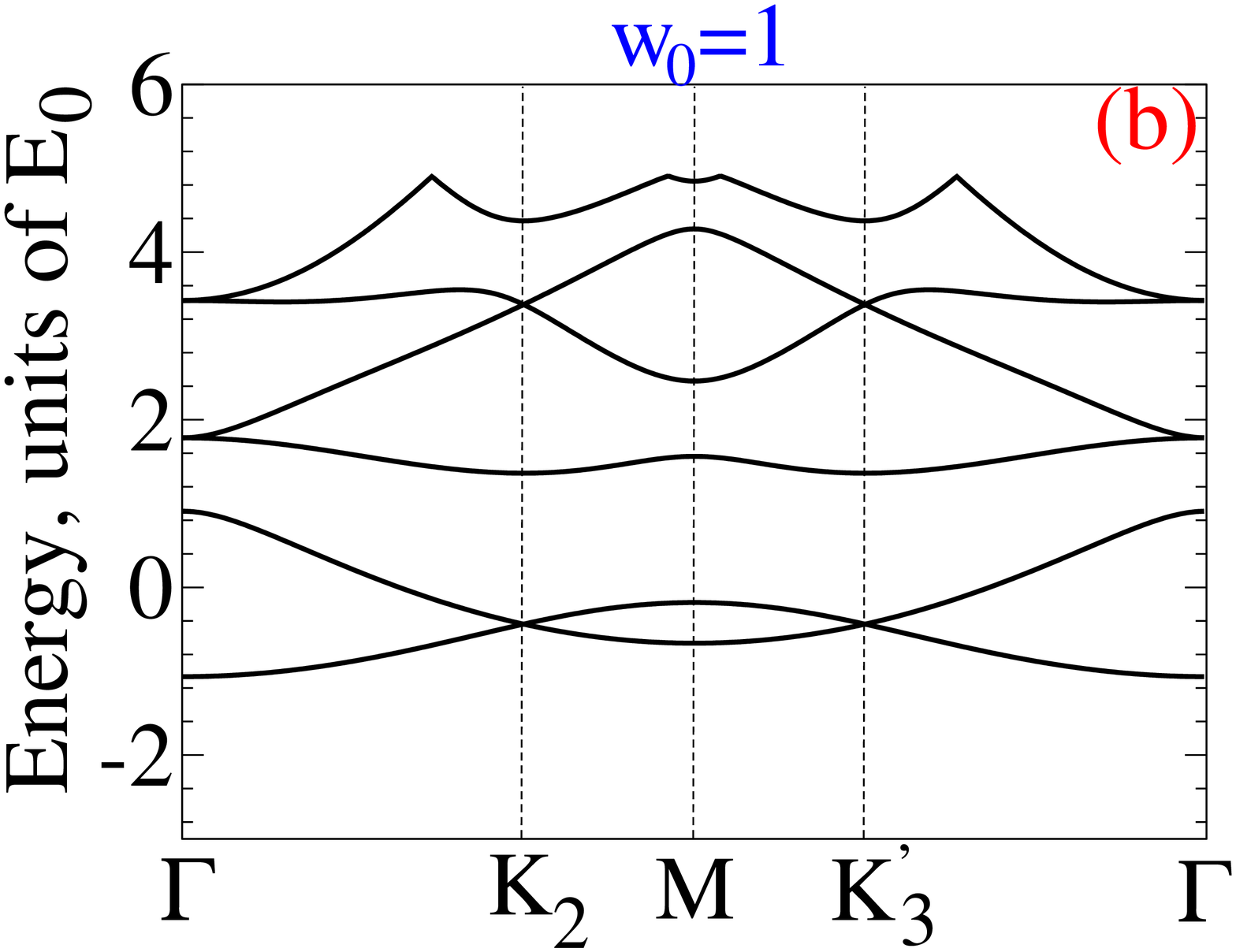}
\includegraphics[width=0.2\textwidth,clip]{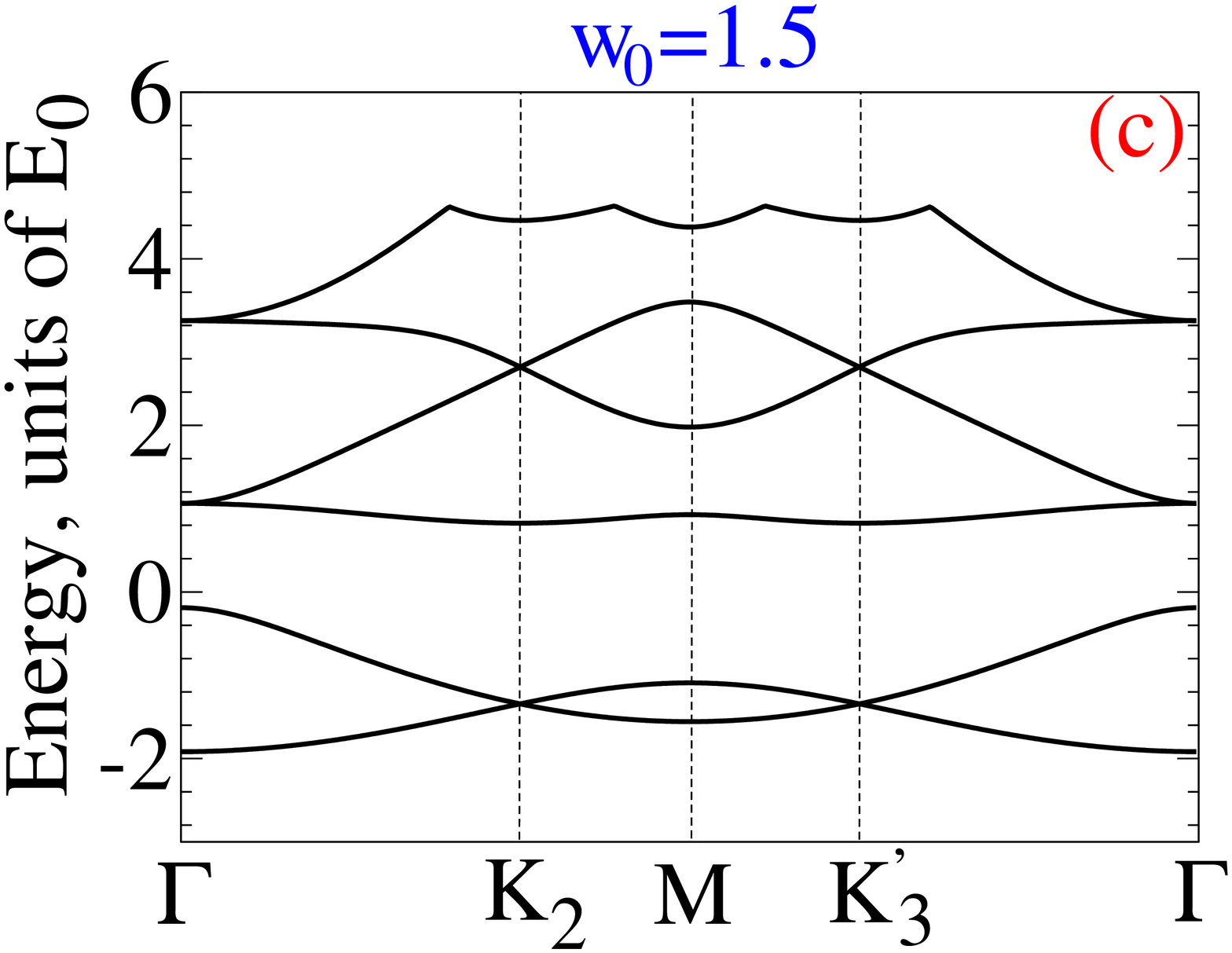}
\includegraphics[width=0.2\textwidth,clip]{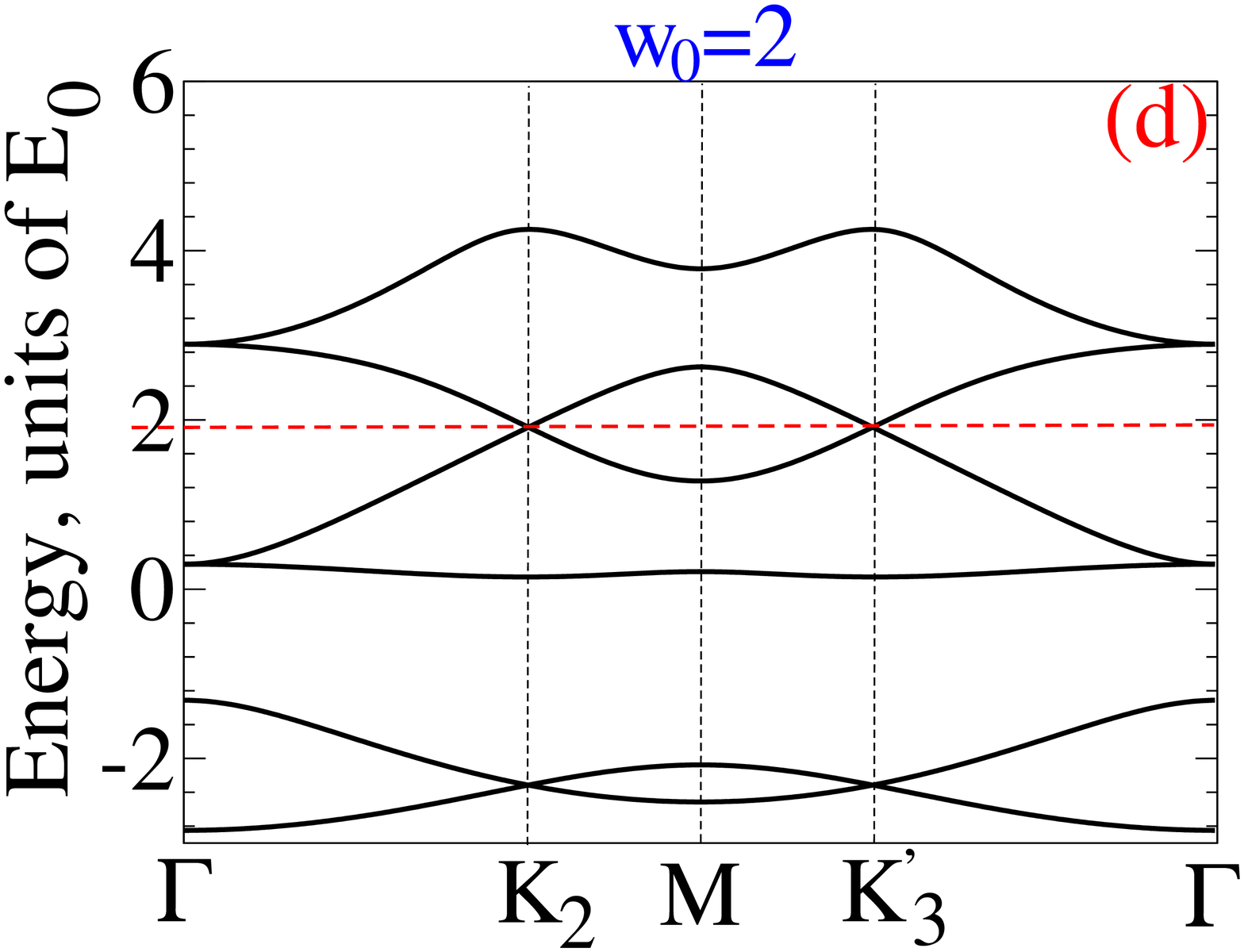}
\caption{Six lowest minibands shown for a particular contour in the
Brillouin zone. This is the case {\it without} account
of the 2DEG self-screening. Different panels correspond to
different values of parameter $\mathrm{w}_0$ defined in Eq.(\ref{e0}).
The first and the second Dirac points are  clearly seen in panels
(c) and (d).
}
\label{F3}
\end{figure}

The miniband structure is determined by the Schr\"odinger equation
\begin{eqnarray}
\label{SE}
\left(\frac{\bm p^2}{2m^*} + U_G(\bm \rho)+U_S(\bm \rho)\right)\psi(\bm \rho)
=E\psi(\bm \rho)\,,
\end{eqnarray}
where $m^*$ is the effective mass  and $U_S$ is due to the self-screening of
2DEG.
In momentum representation the Schr\"odinger equation reads
\begin{eqnarray}
\label{SE1}
\frac{\bm k^2}{2m^*}\psi_{\bm k} + \sum_{{\bm k}'}
\left[U_G({\bm {k-k}}')+U_S(\bm {k-k}')\right]\psi_{{\bm k}'}
=\epsilon_{\bm k}\psi_{\bm k} \ .
\end{eqnarray}
First we disregard the self-screening and set $U_S=0$.
The band structure depends only on the dimensionless parameter $\mathrm{w}_0$
\begin{eqnarray}
\label{e0}
&&\mathrm{w}_0=\frac{U_0}{E_0}\,,\nonumber\\
&&E_0=\frac{K_1^2}{2m^*}=\frac{8\pi^2}{9}\frac{ \hbar^2}{m^*L^2}\ .
\end{eqnarray}
The Fourier component of (\ref{u1}), $U_G({\bm {k-k}}')$, is nonzero only
if ${\bm {k-k}}'= \pm\bm G_i$.
Hence, truncating the summation in (\ref{SE1}) by 20--50 states nearest in
energy one reduces the problem to numerical diagonalization
a matrix of about $100\times 100$ size.
Six lowest calculated bands $\epsilon_{\bm k}$  are shown in Fig.~\ref{F3}
for several values of $\mathrm{w}_0$.
The first and the second Dirac points are well separated and
clearly seen in panels
(c) and (d) of Fig.~\ref{F3}. Higher Dirac points are ``dissolved''
in the dense band spectrum.
Throughout the paper we will assume that the chemical potential is
adjusted to the second Dirac point as it is shown in the panel (d)
Fig.~\ref{F3} by the red dashed line.
The corresponding average electron density  is $n=16/(\sqrt{3}L^2)$.

\subsection{Minibands in the Hartree approximation}
The bands Fig.~\ref{F3} are shown in dimensionless units.
Below in this Section we set
\begin{eqnarray}
\label{L}
&&L=130\,\mathrm{nm},\\
&&E_0=0.59\,\mathrm{meV}. \nonumber
\end{eqnarray}
Here we take $m^*=0.067m_e$.
The average electron density corresponding to the
second Dirac point is $n\approx 0.55 \times 10^{11}\,\mathrm{cm}^{-2}$.
For a usual 2DEG with quadratic dispersion and with dielectric constant
$\epsilon=12$ this density corresponds to $r_s \approx 2.6$.
The average density is  pretty low. A local density is
higher. We will discuss this issue later.

\begin{figure}[ht]
\includegraphics[width=0.2\textwidth,clip]{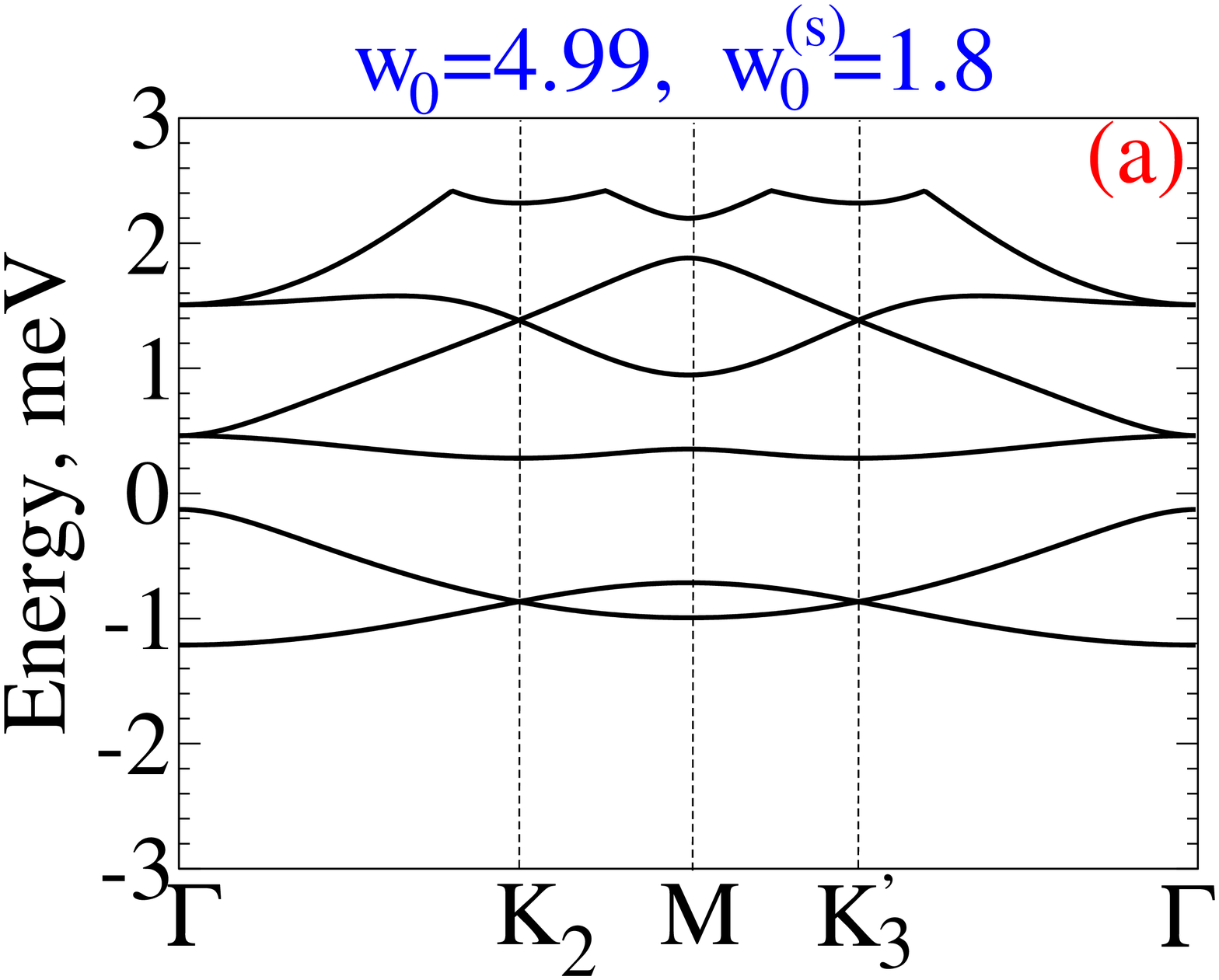}
\includegraphics[width=0.2\textwidth,clip]{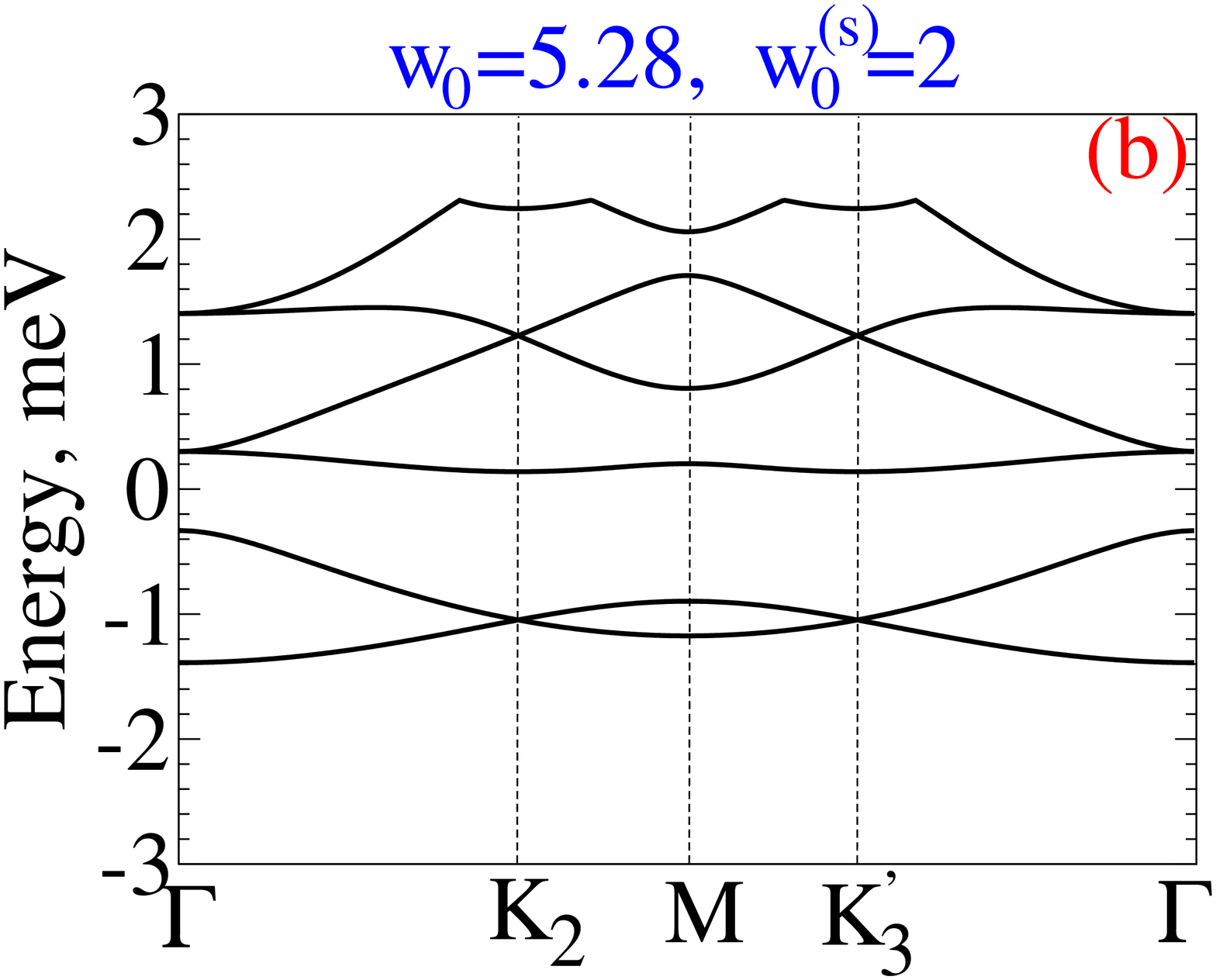}
\includegraphics[width=0.2\textwidth,clip]{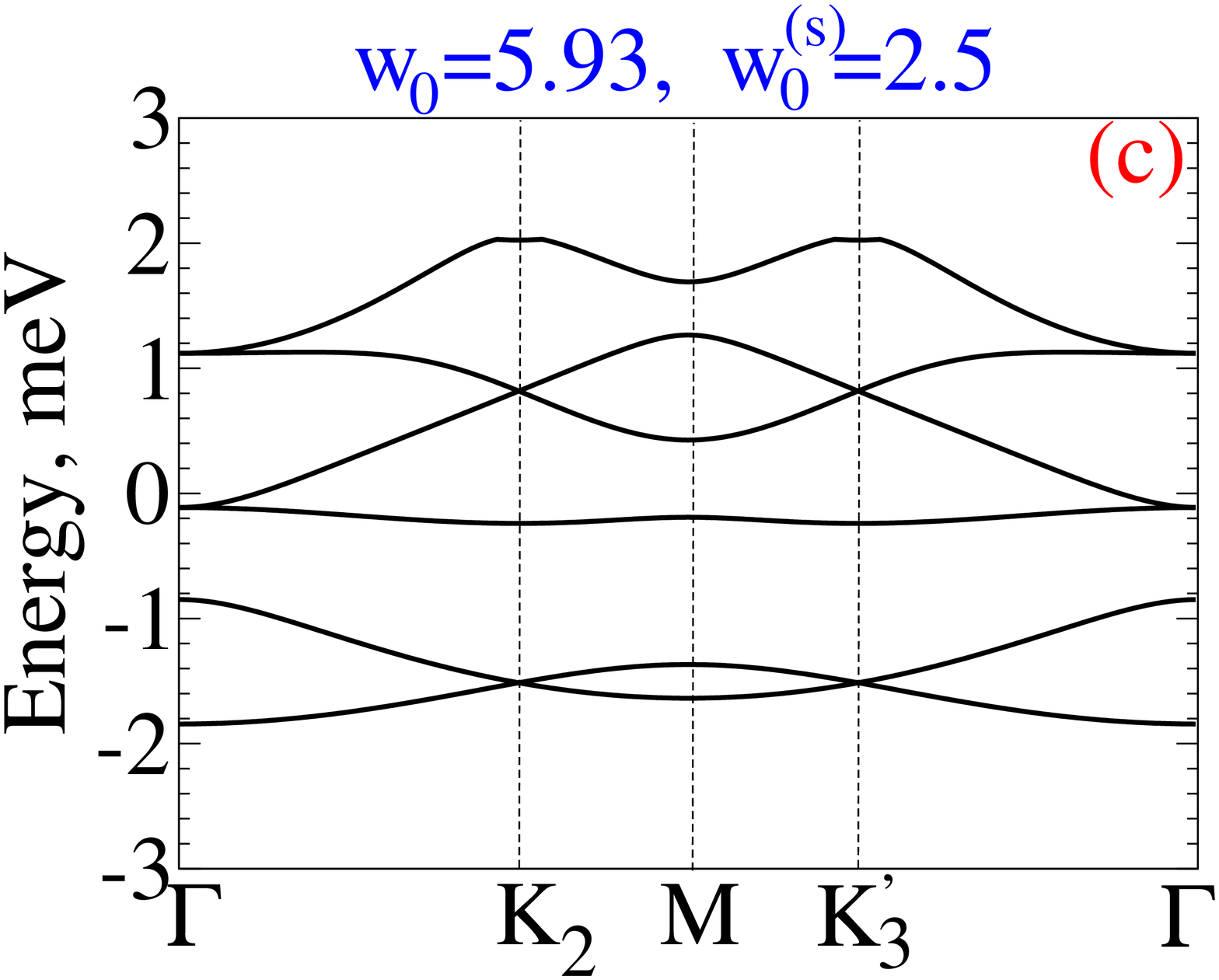}
\includegraphics[width=0.2\textwidth,clip]{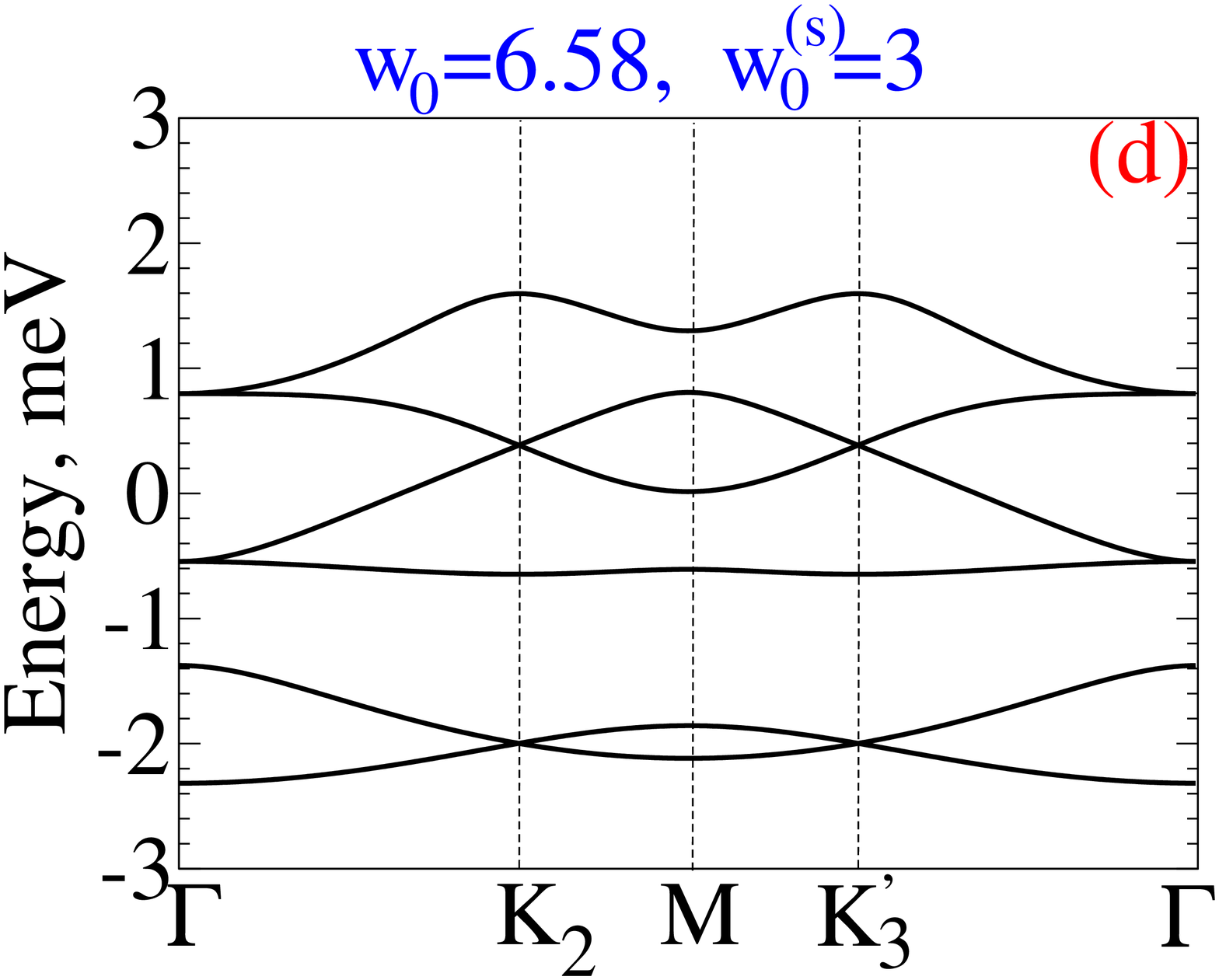}
\caption{Six lowest minibands shown for a particular contour in the
Brillouin zone. This is the case {\it with} account
of the 2DEG self-screening. Different panels correspond to
different values of parameter $\mathrm{w}_0$ defined in Eq.(\ref{e0}).
We also present values of the screened $\mathrm{w}_0^{(s)}$.
We assume $L=130$nm,  so unlike Fig.~\ref{F3} energies are
given in meV.
}
\label{F4}
\end{figure}

According to Fig.~\ref{F3}(d) the Fermi Dirac velocity is
\begin{equation}
\label{vf}
v_F \approx 0.5E_0 L/\hbar \ .
\end{equation}
Hence the ``fine structure constant''
\begin{equation}
\label{al}
\alpha =\frac{e^2/\epsilon}{\hbar v_F} \approx \frac{2e^2/\epsilon}{E_0 L} \to
3.1 \ .
\end{equation}
Here we take the dielectric constant $\epsilon \approx 12$.
Hence, the interaction is stronger than that in
natural suspended graphene where $\alpha \approx 1{-}2$, see
Ref.~\onlinecite{CastroNeto09}.
To  account for the 2DEG self-screening $U_S$  we use the Hartree
approximation and solve Eq.(\ref{SE1}) iteratively.
We start from the noninteracting solution described in two previous
paragraphs. Using this solution we calculate the electron density
\begin{eqnarray}
\label{den}
 n(\bm \rho)=2\sum_{{\bm k}, \epsilon_{\bm k} < \mu}|\psi_{\bm k}(\bm \rho)|^2\,.
\end{eqnarray}
Then we Fourier transform the density  numerically,
$n(\bm \rho) \to  n(\bm q)$, and hence find $U_S(\bm q)$
\begin{eqnarray}
\label{usk}
U_S(\bm q)=\frac{2\pi e^2/\epsilon}{q}n(\bm q) \ .
\end{eqnarray}
At the next iteration we substitute (\ref{usk}) in Eq. (\ref{SE1}),
find new wave functions and dispersions, calculate the
electron density (\ref{den}) and hence again calculate the
screening potential (\ref{usk}).
The iteration procedure converges after several iterations.
When  doing the iterations we keep the chemical potential adjusted
to the  second Dirac point, so we keep the average electron
density unchanged. The self-screening potential $U_S$ has two effects.
(i) $U_S$ partially screens the first harmonic in the gate potential
(\ref{u1}), $U_0 \to U_0^{(s)} < U_0$.
(ii) $U_S$ generates higher harmonics in the effective potential.
Six lowest bands $\epsilon_{\bm k}$  calculated with
account of screening are shown in Fig.~\ref{F4} for several values of $\mathrm{w}_0$.
In the panels we present values of $\mathrm{w}_0$ and also values of the screened
constant $\mathrm{w}_0^{(s)}=U_0^{(s)}/E_0$.
The spacial distribution of 2DEG electron density at parameters corresponding
to Fig.\ref{F4}(d) is illustrated in Fig.\ref{F5}.
Here we would like to stress two points; (i) The electron density,
Fig.\ref{F5}(a), is connected, so this is the anti-dot regime,
(ii) while the average electron density is $0.55\times10^{11}\,\mathrm{cm}^{-2}$, the
density along the ``connected manifold'' is larger than $10^{11}\,\mathrm{cm}^{-2}$.
The value of the external potential modulation corresponding
to Fig.\ref{F4}(d) is $\mathrm{w}_0=6.58$. Using Eqs.(\ref{u1}),(\ref{e0}),(\ref{L})
we can find that the corresponding value of the gate voltages is
($L=130$\,nm, $z=37+8=45$\,nm)
\begin{equation}
\label{vg}
\VTG-\VPG=-0.7\,\mathrm{V}.
\end{equation}

\subsection{Minibands in the Poisson-Hartree approximation}
The calculation described above has two inaccuracies, (i) Eq.(\ref{usk})
implies that the gate screening
of the 2DEG self-screening (screening of screening)
is not taken into account, (ii) we assume that
the width of the 2DEG quantum well is very small $W=0$.
To fix these problems we perform an alternative, more involved and more
accurate calculation  of the  miniband structure.
We call this calculation the Poisson-Hartree calculation.
We assume the following geometrical parameters, the lattice spacing $L=130$\,nm,
the perforation diameter $D=60$\,nm,
the distance from gates to 2DEG $z=37$\,nm, the width of the rectangular
quantum well $W=16$\,nm. The wave function along the $z$-direction
is the lowest standing wave in the well,
$\psi_z(z)=\sqrt{\frac{2}{W}}\sin(\pi (z-z_0)/W)$.
Hence the total wave function is
$\psi({\bm r})=\psi_{\bm k}({\bm \rho})\,\psi_z(z)$.
\begin{figure}[ht]
\includegraphics[width=0.23\textwidth,clip]{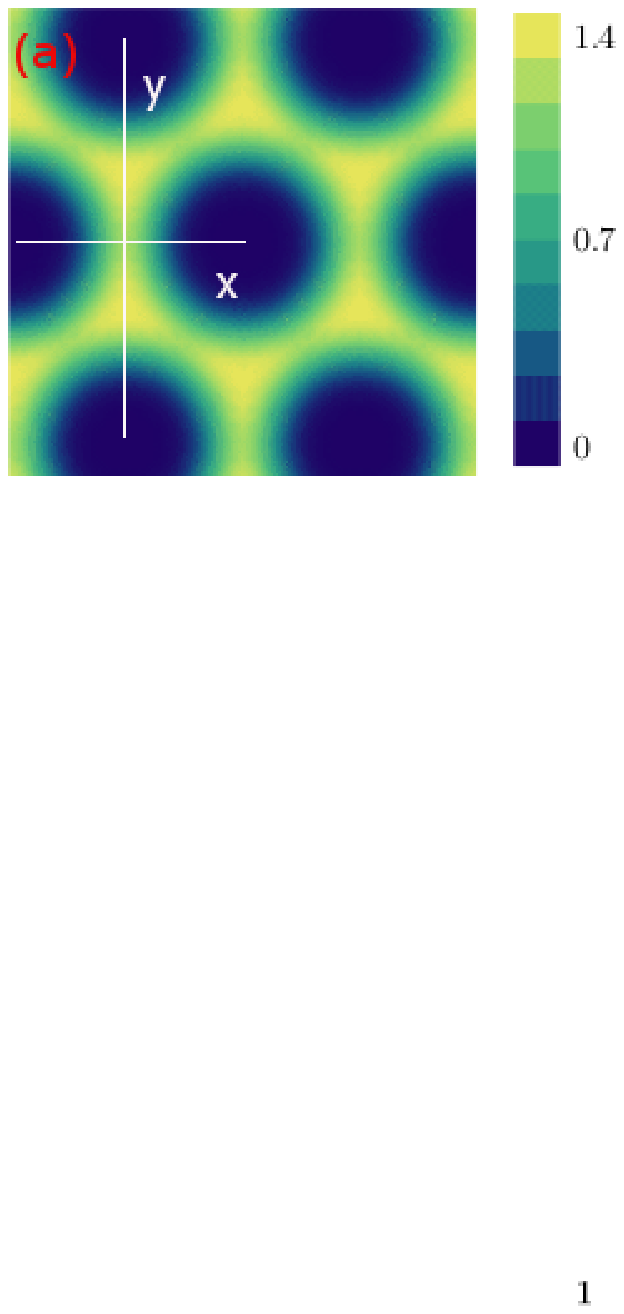}
\includegraphics[width=0.23\textwidth,clip]{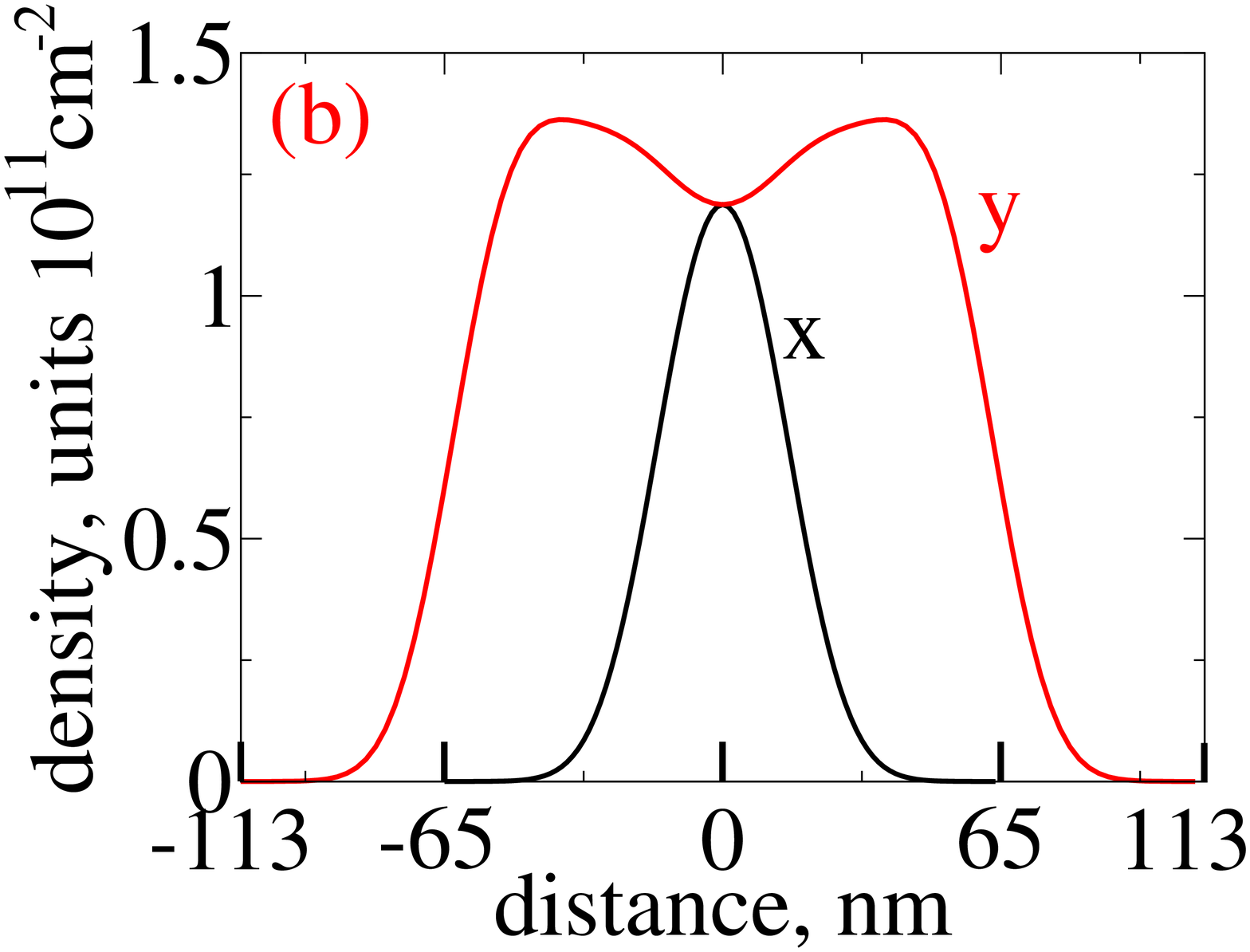}
\caption{Panel~(a) presents the map of electron density in units
of $10^{11}$\,cm$^{-2}$. Panel~(b) presents plots of electron density
along $x$- and $y$-directions indicated in panel~(a).
Parameters of the potential correspond to that presented in Fig.~\ref{F4}(d).
}
\label{F5}
\end{figure}
Instead of Eq.~\eqref{den} the 2DEG electron density is
\begin{eqnarray}
\label{den1}
 n(\bm r)=2\sum_{{\bm k}, \epsilon_{\bm k} < \mu}|\psi_{\bm k}(\bm \rho)|^2
|\psi_z(z)|^2 \,.
\end{eqnarray}
\begin{figure}[hb]
\includegraphics[width=0.23\textwidth,clip]{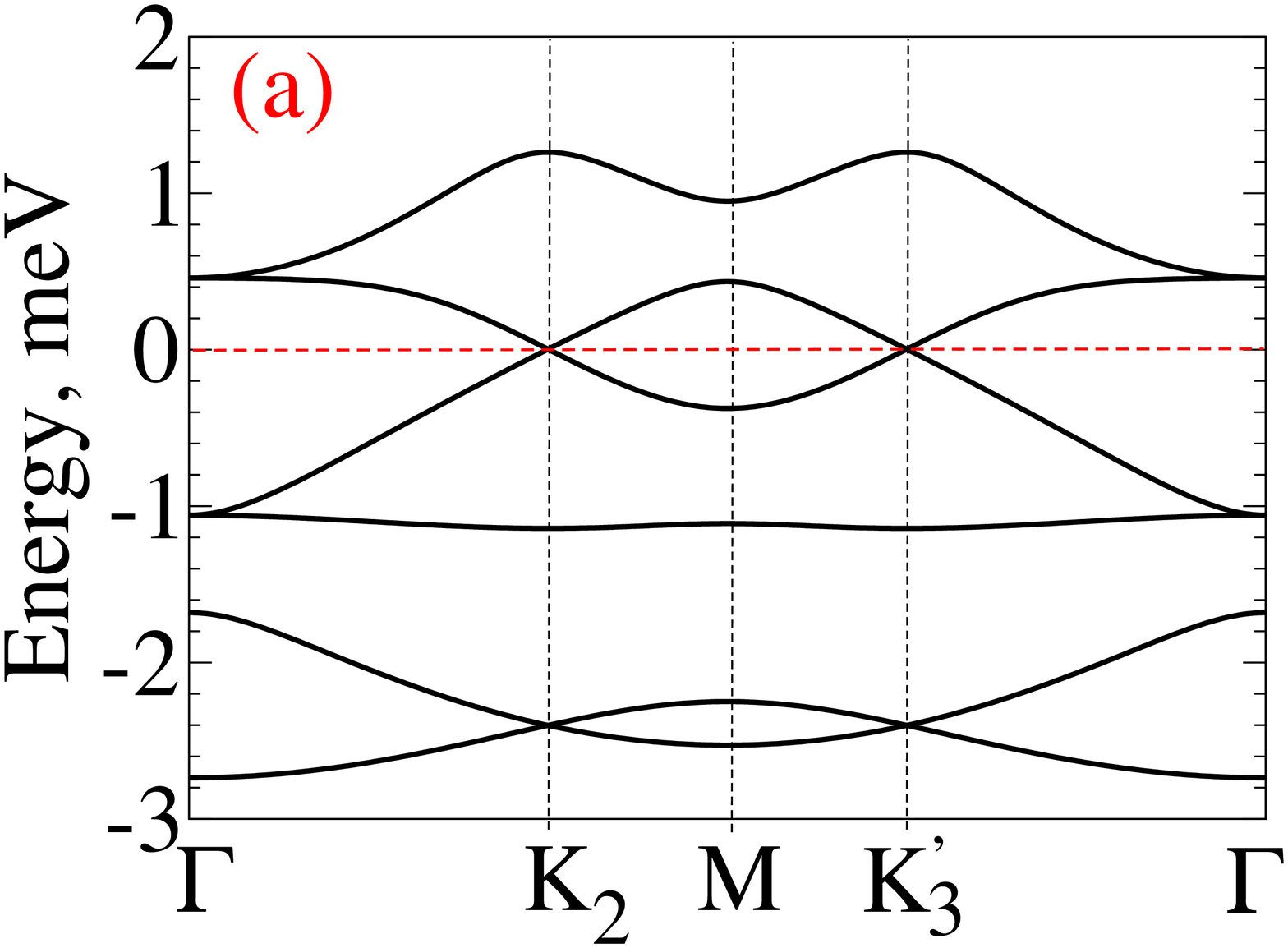}
\includegraphics[width=0.23\textwidth,clip]{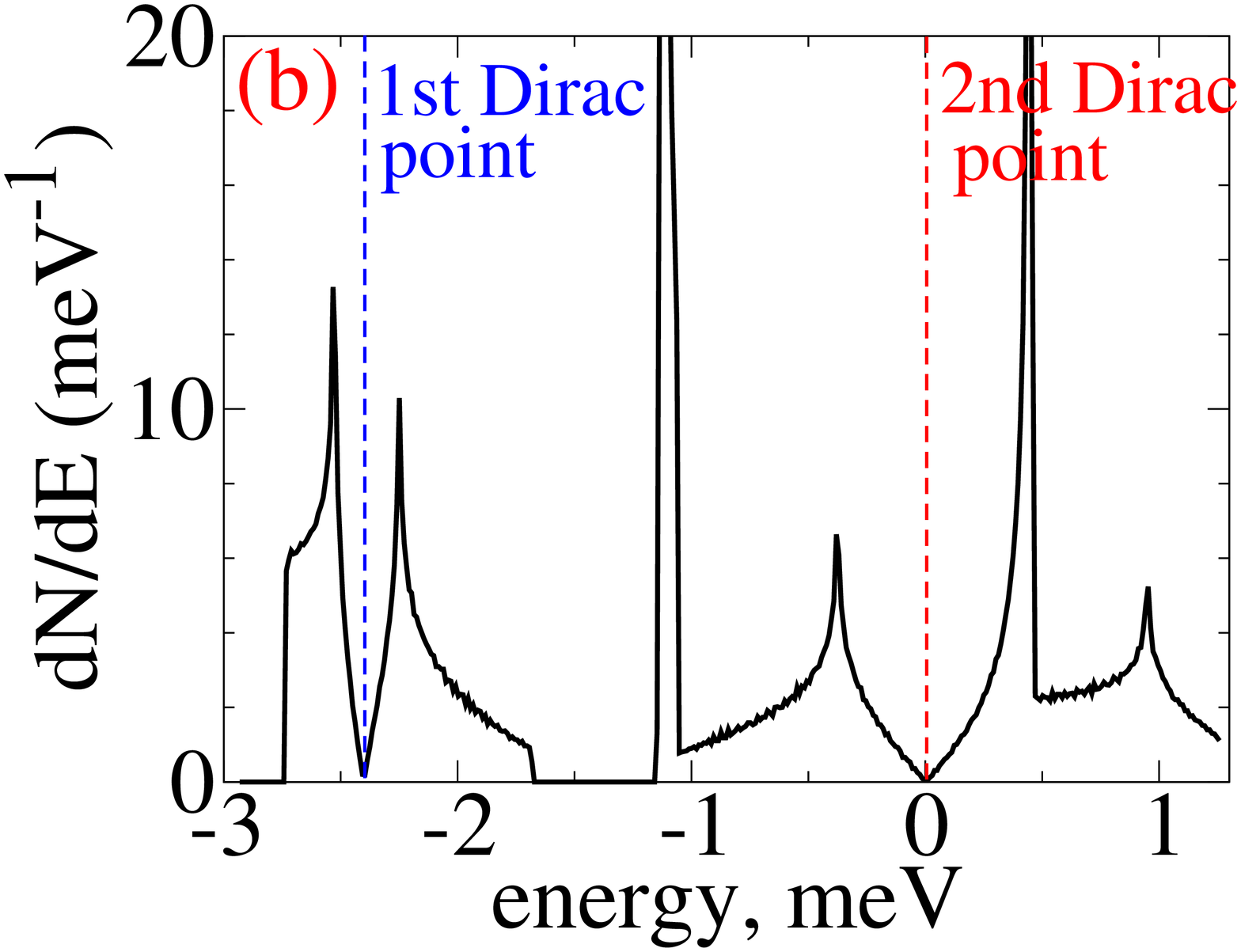}
\includegraphics[width=0.23\textwidth,clip]{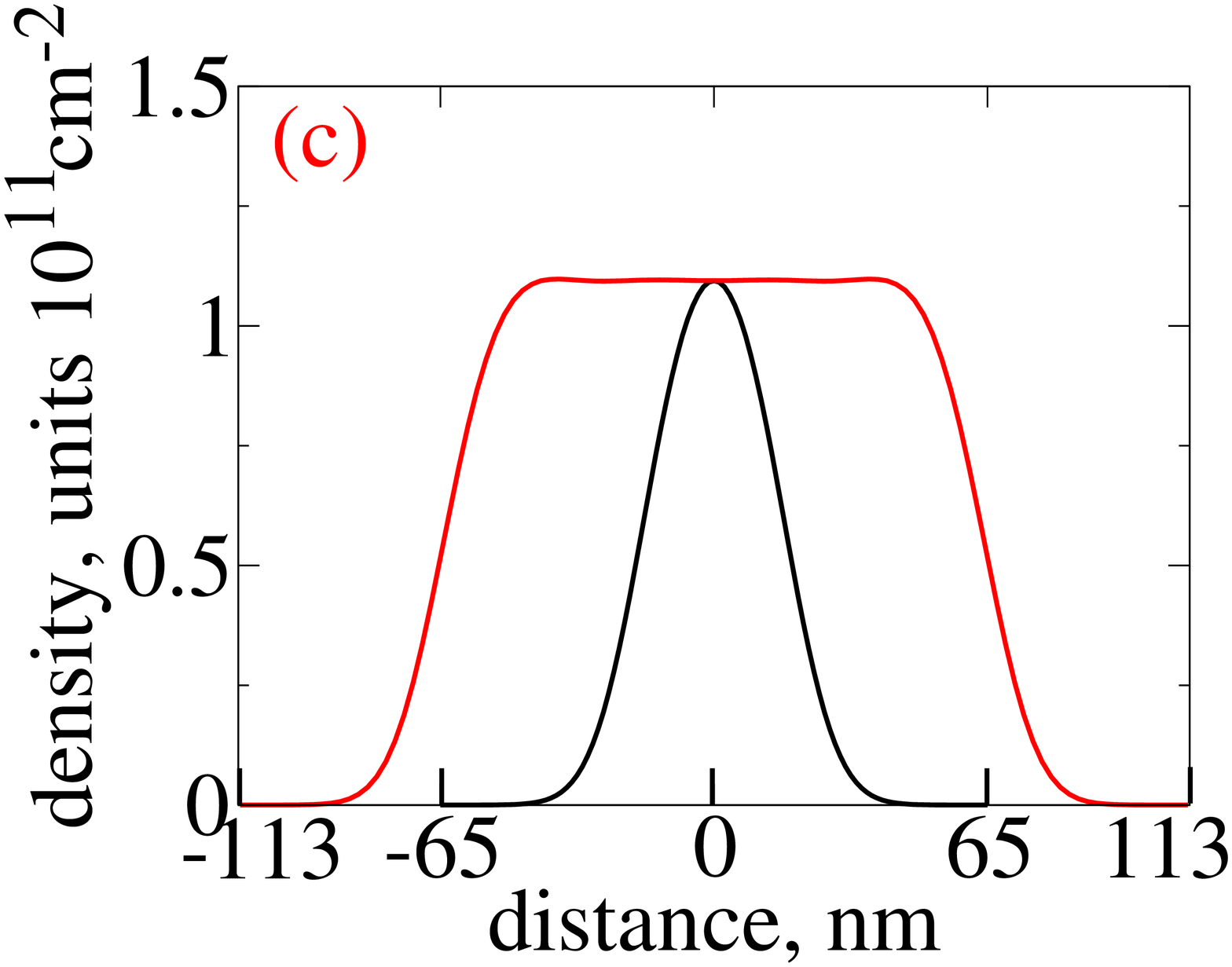}
\includegraphics[width=0.23\textwidth,clip]{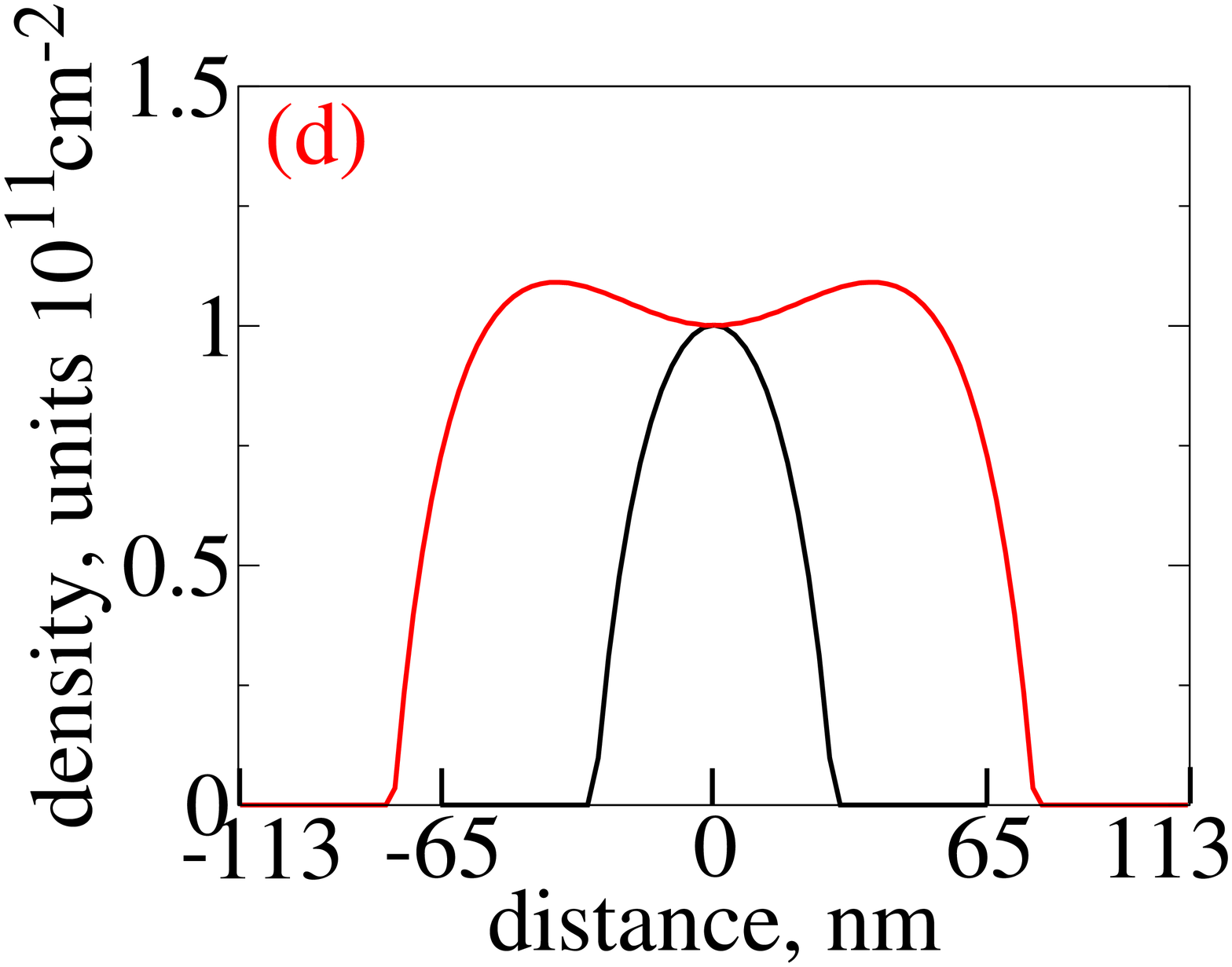}
\caption{Miniband structure properties for  the following gate voltages,
$\VTG=0.4$\,V and $\VPG=0.84$\,V and the following
geometrical parameters, $L=130$\,nm, $D=60$\,nm, $z=37$\,nm,  $W=16$\,nm.
Panels (a), (b), (c) show results of the Poisson-Hartree calculation:
(a) six lowest minibands for a particular contour in the
Brillouin zone; (b) the density of states per the superlattice unit cell;
(c) plots of electron density along the same directions as that
in Fig.~\ref{F5}.
Red dashed lines in panels (a) and (b) indicate the chemical potential.
Panel (d) shows  plots of electron density similar to that in
panel (c), but obtained within the Poisson-TF calculation.
}
\label{F6}
\end{figure}
To find the electrostatic potential we solve Poisson equation,
the gate voltages are imposed via boundary conditions at the metallic
surfaces. With this potential we solve 2D  Schr\"odinger Eq.~\eqref{SE},
and find wave functions $\psi_{\bm k}(\bm \rho)$. Then we substitute this in
Eq.~\eqref{den1} and iterate the procedure.
This is the same Hartree procedure, but it accounts for finite $W$ and
fully accounts for all screenings.
Results of this calculation at $\VTG=0.4$\,V and $\VPG=0.84$\,V
 are presented in panels (a),(b), and (c) of  Fig.~\ref{F6}.
The gate voltages are adjusted to tune the chemical potential to
the 2nd Dirac point, and to provide a sufficient modulation of
the superlattice potential.
The miniband structure Fig.\ref{F6}(a) and the electron density
Fig.\ref{F6}(c) are pretty close to those obtained within the Hartree method
and shown in Fig.\ref{F4}(d) and Fig.\ref{F5}.
Fig.\ref{F4}(d) displays the situation close to the optimal one for the
artificial graphene. We deliberately adjusted gate voltages to values
\begin{equation}
\label{vg1}
\VTG-\VPG=0.4-0.84=-0.44\,\mathrm{V},
\end{equation}
which approximately reproduce parameters corresponding to Fig.~\ref{F4}(d).
The major difference between the result of the Hartree calculation
and the result of the Poisson-Hartree calculation is
the difference between Eq.(\ref{vg}) and Eq.~\eqref{vg1}.
The Hartree method overestimates the value of $\VTG-\VPG$ because
screening by gates is ignored in this consideration.
As soon as  $\VTG-\VPG$ is adjusted,
all other results between the two methods are very close.

\subsection{Minibands in the Poisson-Thomas-Fermi approximation}
The analysis presented above is absolutely sufficient  for an ideal
superlattice. Unfortunately in a disordered system we cannot perform such an
analysis because there is no band structure and there is no a
well defined quasimomentum. Therefore, for a disordered  system we will
use the Poisson-Thomas-Fermi (Poisson-TF)  method
(see, for example, \onlinecite{pyshkin,davies,modeling}).
Here we want to check how the method works for the perfectly periodic
system.
Idea of the method is very simple, instead of solving
Schr\"odinger equation (\ref{SE}) and then using
(\ref{den}) to find the electron density, we use the
local 2D Thomas-Fermi approximation to determine the density
\begin{equation}
\label{TF1}
n({\bm \rho})=\frac{m^*}{\pi \hbar^2}
\left[\mu_{\mathrm{TF}}-U_{\mathrm{eff}}({\bm \rho})\right].
\end{equation}
Here $\mu_{\mathrm{TF}}$ is the effective chemical potential,
and $U_{\mathrm{eff}}$ is the effective self-consistent potential.
If  expression in  brackets in Eq.(\ref{TF1}) is negative, we set
$n({\bm \rho})=0$.
The effective chemical potential is determined by the equation
\begin{equation}
\label{mtf}
\int n({\bm \rho})d^2\rho=n \ ,
\end{equation}
where the integration is performed over the lattice unit cell,
and $n$ is the average electron density, $n=8$ for the second Dirac point.
Using $n({\bm \rho})$ from Eq.(\ref{TF1}) we solve the Poisson equation
with appropriate boundary conditions at metallic gates, determine
$U_{\mathrm{eff}}({\bm \rho})$, and repeat the procedure iteratively.
After completion of the iterative procedure we solve the
Schr\"odinger equation, find the miniband structure, density of states, etc.
By changing gate voltages we match the position of the second Dirac point
to zero energy.
The band structure and the density of states determined by this method
are  practically identical to that found within the Poisson-Hartree method
and shown in panels (a),(b) of Fig.~\ref{F6}.
The electron density, Fig.~\ref{F6}(d), is slightly different, though
it is very close to the
Poisson-Hartree density shown in panel (c).
The main lesson we learn from this comparison is that we can rely on the
Poisson-TF method for studies of disordered systems.

\subsection{Recursive Green's functions and conductance}

\begin{figure}[hb]
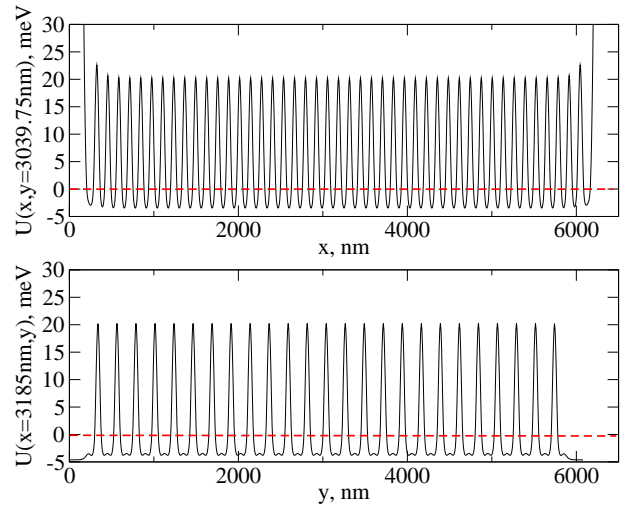

\includegraphics[width=0.45\textwidth,clip]{fig07a_ux.eps}
\includegraphics[width=0.45\textwidth,clip]{fig07b_uy.eps}
\caption{Self-consistent potential energy for a perfect (no disorder)
$6\mum\times6\mum$ artificial graphene. The energy is plotted
along the $x$- and the $y$-directions, see also Fig.\ref{tri}.
The chemical potential shown by the red dashed line is zero.
}
\label{uxy}
\end{figure}

\begin{figure}[ht]
\includegraphics[width=0.23\textwidth,clip]{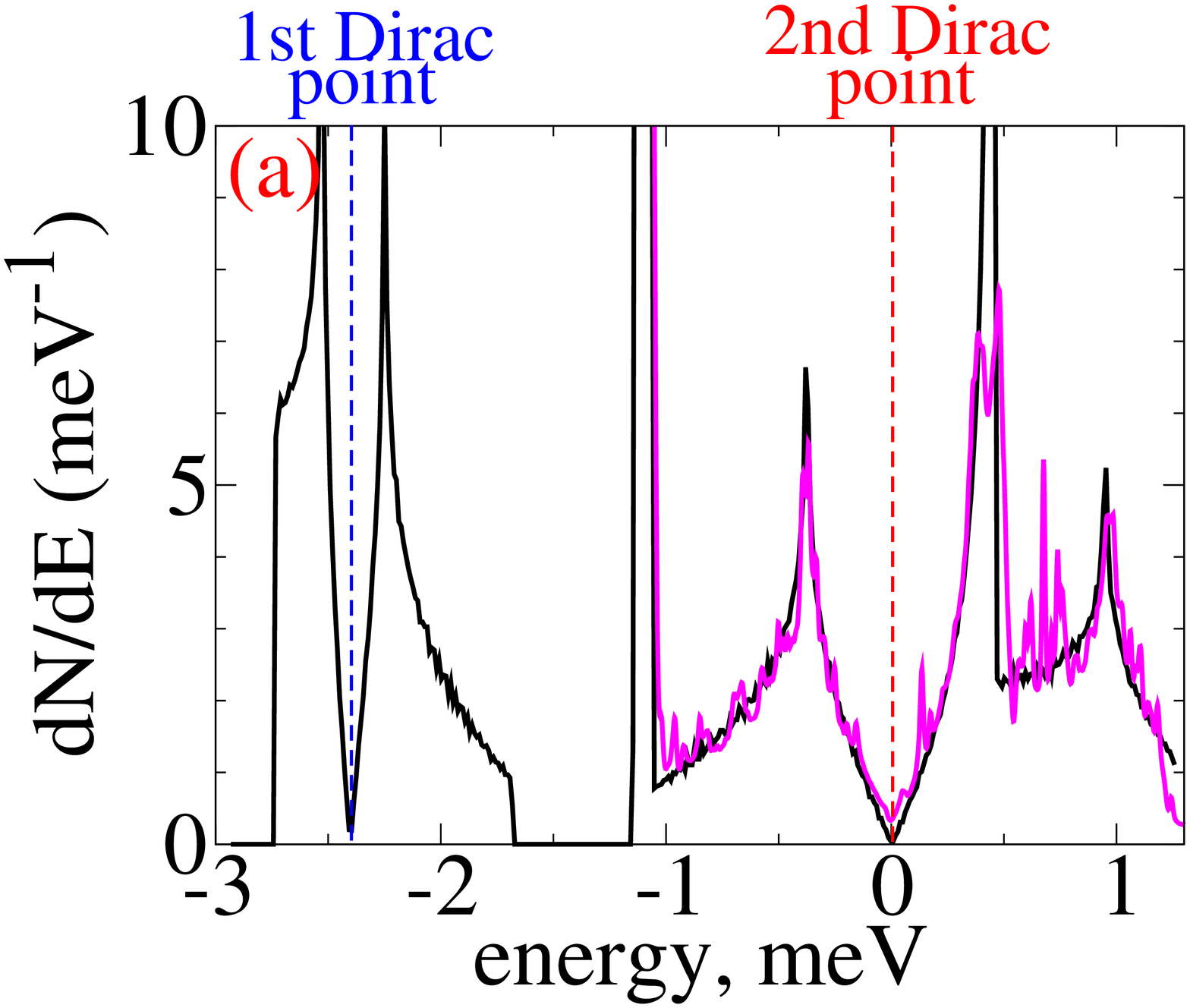}
\includegraphics[width=0.23\textwidth,clip]{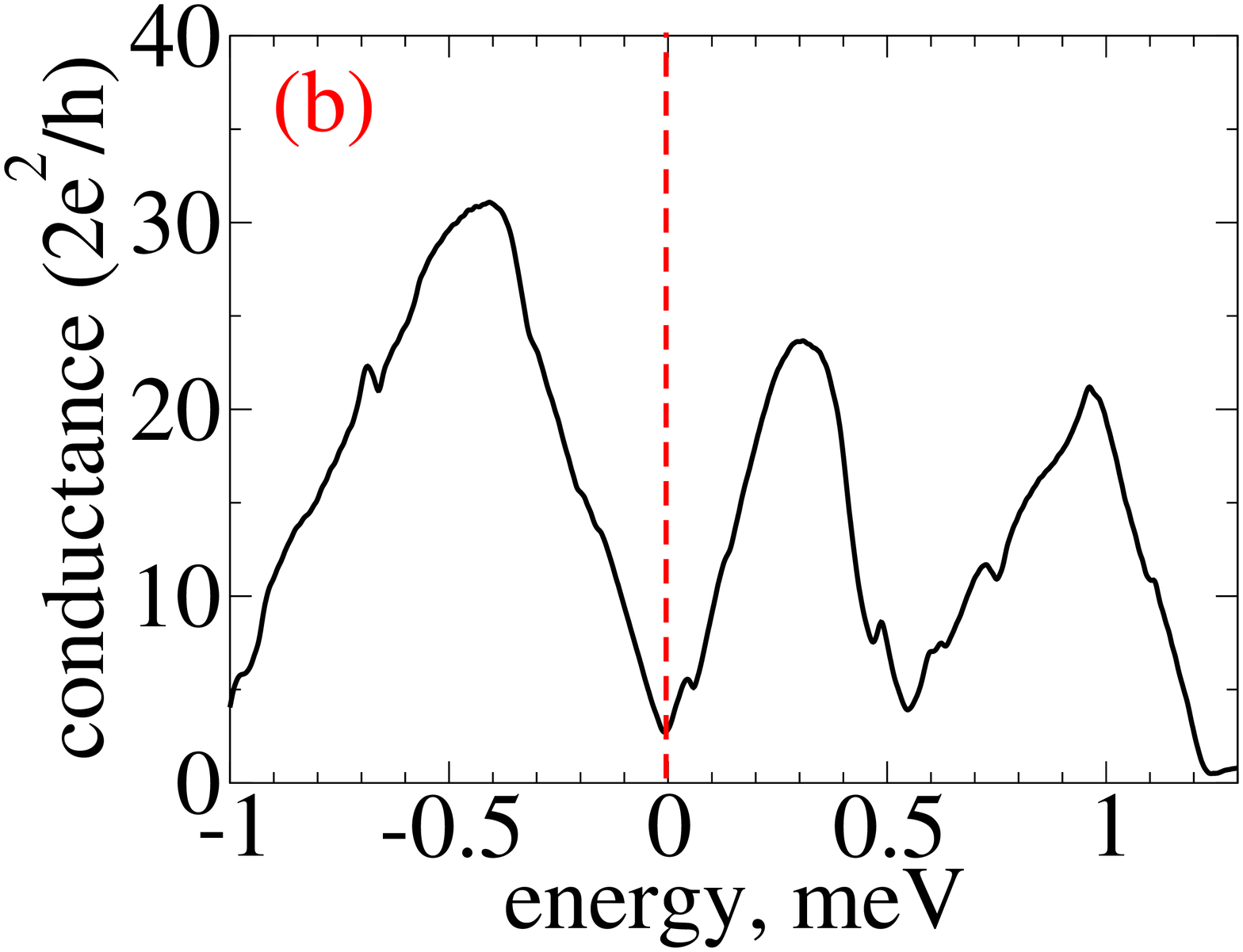}
\caption{(a) The density of states per the  superlattice unit cell.
(b)~The conductance of a perfect (no disorder) artificial graphene
sample of $6\mum\times6\mum$ size.
The noisy magenta line in panel~(a) shows the density of states
calculated by
the  recursive Green's function method. The black solid line in the same panel,
identical to that in Fig.~\ref{F6}(b), is the density of states calculated
via the band dispersion.
}
\label{dos}
\end{figure}

As soon as the effective potential is calculated by  the Poisson-TF method we
can calculate conductance of the artificial graphene ``sample.''
 To do so we use the recursive Green's function method~\cite{Ando1991,Cresti2003}
which is efficient for calculation of single particle \cite{pyshkin}.
We consider a ``sample'' of approximately $6\mum\times6\mum$ size.
In the $x$-direction the  size is limited by the width of the top gate. In the $y$-direction the sample is connected to the metallic leads. The self-consistent
potential plotted along the $x$- and $y$-directions is shown in Fig.~\ref{uxy}.
This is the case of M-orientation of anti-dot lattice with zigzag-configuration at the entries and with armchair-configuration along channel edges.
Electric current flows in the $y$-direction.
To calculate Green's functions we use effective potential interpolated on
a coarse square grid (638$\times$574 sites) with 10\,nm steps in $x$- and
$y$-directions. The original  $U_{\mathrm{eff}}$, obtained within the Poisson-TF
approach, is more detailed than the coarse grid we use for Green's functions.
We use the coarse grid since Green's functions are computationally
more intensive.
The density of states calculated by the recursive Green's function
method is plotted in panel (a) of Fig.\ref{dos} by the ``noisy''
magenta line.
The density of states is very close to that calculated from the band dispersion
and shown by the solid black line in the same panel.
The ``noise'' in the density of states calculated by the recursive Green's
function method is due to the finite size of the ``sample.''

The calculated conductance is presented in panel (b) of  Fig.~\ref{dos}.
Interestingly, at the Dirac point, $\epsilon=0$, the conductance
does not dive to zero, $G(0)\approx 2\times(2e^2/h)$.
We have checked that this is due to the two edge states located at $x\approx 350$~nm
and $x\approx 6000$~nm. The edge states can be removed by a small
variation of the gate potentials (for example, there are no such states
at $\VTG=0.48$~V,
$\VPG=0.835$~V, in this case $G(0)=0$).
The edge states are not topologically protected, therefore even a small
disorder will remove/influence the states.
Now we are fully armed to consider the effect of disorder
or in other words to consider an imperfect gate structure.\\

\section{Influence of disorder\label{Section2}}
\subsection{Types of disorder}
There are three different types of the gate disorder:
(i) the area disorder,
(ii) the position disorder,
(iii) the shape disorder.
In the case of the area disorder the radius of perforation, see Fig.\ref{tri},
is randomly changed
\begin{equation}
\label{ad}
R=\frac{D}{2} \to R=\frac{D}{2}+ \delta_A(0.5-{\cal R})\ ,
\end{equation}
where ${\cal R}$ is a random variable distributed uniformly in the interval
$0< {\cal R}< 1$. In this case the area $A$ of the perforated circle is randomly
changed  with the following rms variation
\begin{equation}
\label{arms}
\frac{\delta A_{rms}}{A}\approx \frac{2}{\sqrt{3}}\frac{\delta_A}{D} \ .
\end{equation}
We assume that $\delta_A \ll D$, therefore Eq.~(\ref{ad}) is equivalent to a
ring with random potential added to the regular gate potential
\begin{equation}
\label{ad1}
\delta V_A({\bm r})=(\VTG-\VPG)\delta_A(0.5-{\cal R})\delta(r-D/2)\ .
\end{equation}
Here $\delta(x)$ is the Dirac $\delta$-function.

In the case of the position disorder the position of the perforation
circle center is randomly changed
\begin{equation}
\label{pd}
{\bm r} \to {\bm r}+{\bm n}\delta_P{\cal R}\ ,
\end{equation}
where ${\bm n}$ is a unit vector with a random orientation in the plane.
Again this is equivalent to a  ring with random potential added to the regular gate potential
\begin{equation}
\label{pd1}
\delta V_P({\bm r})=(\VTG-\VPG)\delta_P {\cal R} \cos\theta \delta(r-R),
\end{equation}
where $\theta$ is the angle between ${\bm n}$ and ${\bm r}$.

In the case of the shape disorder the shape of the perforation
circle is randomly changed, say circular to elliptic.
This is also equivalent to a  ring with the following
random potential added to the regular gate potential
\begin{equation}
\label{sd1}
\delta V_S({\bm r})=(\VTG-\VPG)\delta_S{\cal R} (2\cos^2\theta-1) \delta(r-R),
\end{equation}
where $\theta$ is the angle between ${\bm r}$ and the axis of the ellipse.

It is easy to check that 2D Fourier transforms of the disorder potentials
(\ref{ad1}),(\ref{pd1}),(\ref{sd1}) behave differently at small momenta,
$q \to 0$,
\begin{eqnarray}
\label{asim}
&&\delta V_A({\bm q}) \propto q^0=1\nonumber\\
&&\delta V_P({\bm q}) \propto q\nonumber\\
&&\delta V_S({\bm q}) \propto q^2 \ .
\end{eqnarray}
The small $q$ components are the most  important ones away from the gate plane
since due to the Laplace equation they decay with $z$ as
\begin{equation}
\label{dec}
\delta V(q,z)\propto \delta V(q,0)e^{-qz} \ .
\end{equation}
From here we immediately conclude that assuming
$\delta_A \sim \delta_P \sim \delta_S$ the effect of the area
disorder is the most important one in the 2DEG plane.
The effect of the position disorder on 2DEG is less important and the
effect of the shape disorder is even smaller.
We have checked this analytical conclusion by direct numerical simulations.
Below we consider only the most dangerous area disorder.

\begin{figure}[ht]
\includegraphics[width=0.4\textwidth,clip]{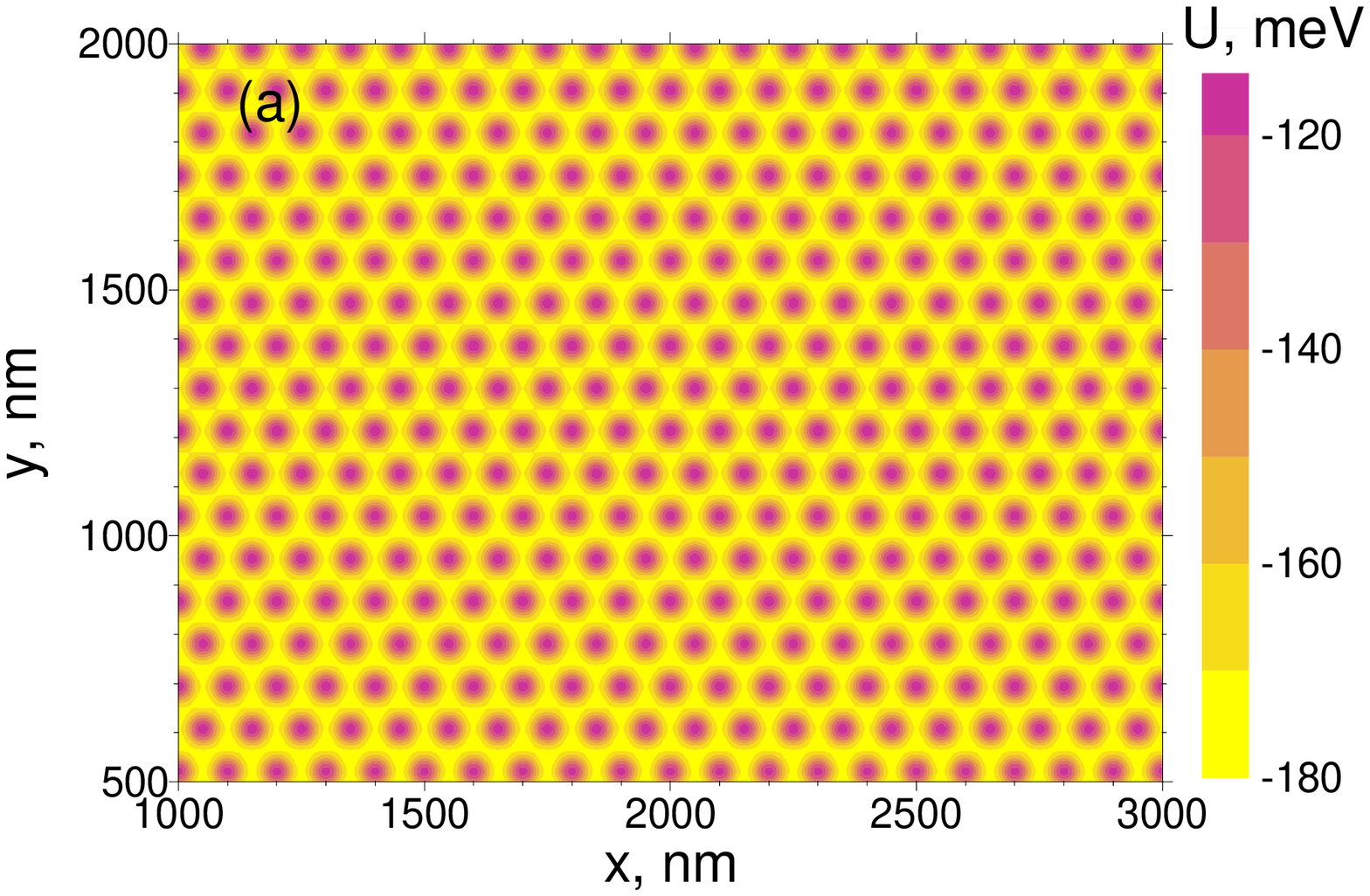}
\includegraphics[width=0.4\textwidth,clip]{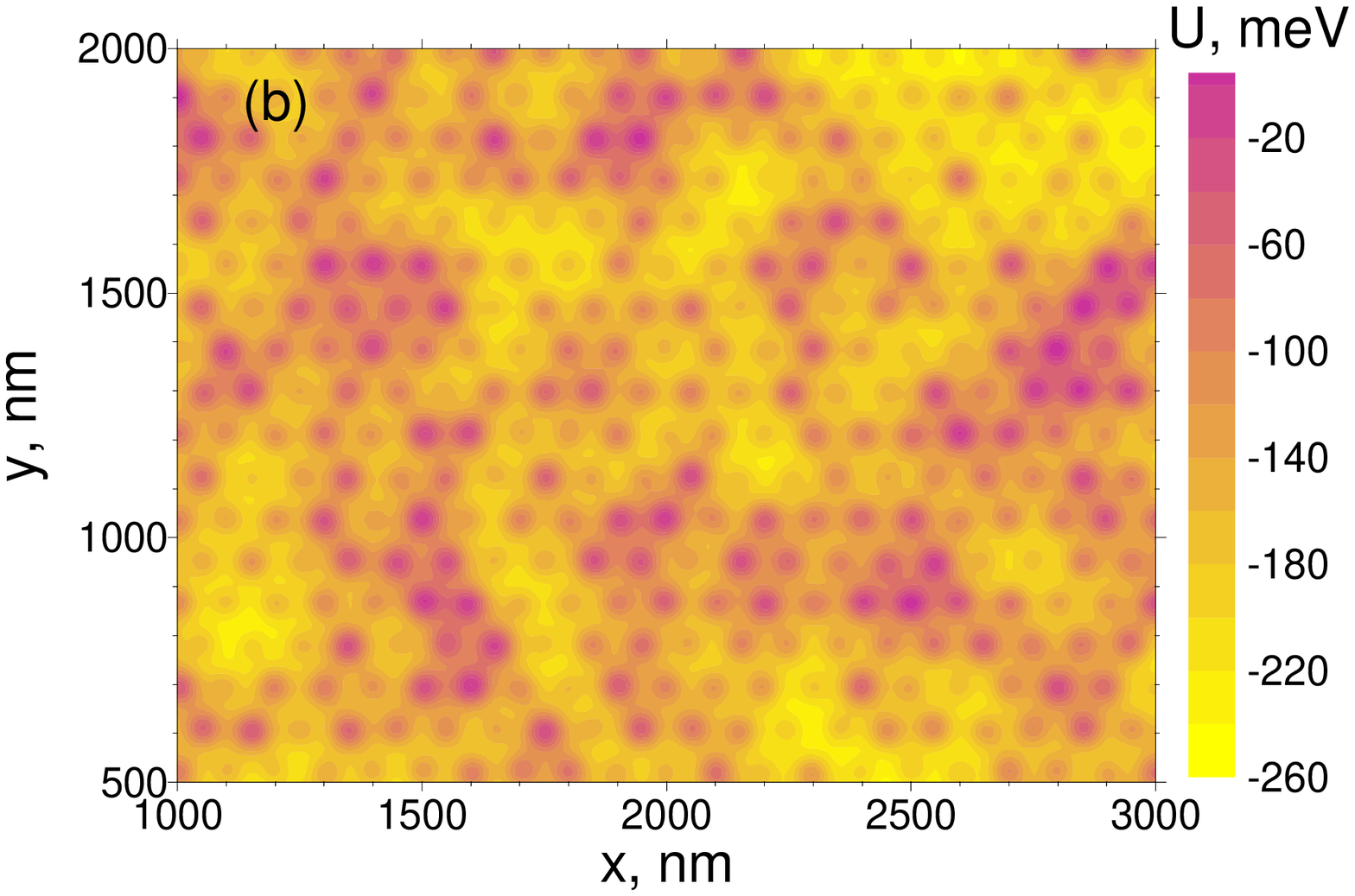}
\caption{Maps of the potential energy in the 2DEG plane
for particular realizations of the gate disorder.
The heterostructure parameters are
$L=100$\,{nm}, $D=50$\,{nm}, $z=50$\,{nm}, $W=16$\,{nm},
$\VPG-\VTG=4$\,V.
Panel~(a) corresponds to the perfect gate structure, $\delta_A=0$.
Panel~(b) accounts for the area disorder parameter $\delta_A=2.5$~nm,
which is equivalent to 5\% rms variation of the area of gate system anti-dots.
}
\label{UxyId100}
\end{figure}

\subsection{Sensitivity to parameters and acceptable design}
First we demonstrate dramatic sensitivity to parameters of the system.
To do so, we do not solve selfconsistently Poisson-TF equations.
Instead we just solve electrostatic Laplace equation.
In doing so we keep in mind that the gate system must support
artificial graphene. This implies that if we vary the lattice spacing
$L$ the gate voltage must scale as $\VPG-\VTG\propto 1/L^2$,
see Eq. (\ref{e0}), and, if vary
the distance $z$ to 2DEG  the gate voltage must scale as
$\VPG-\VTG\propto \exp\left\{\frac{4\pi z}{\sqrt{3}L}\right\}$,
see Eq. (\ref{u1}).
In the panel (a) of Fig.\ref{UxyId100} we present map of the electrostatic
potential of the perfect (no disorder) gate system with following parameters
\begin{eqnarray}
\label{par1}
&&L=100\,\mathrm{nm}, \ \ D=50\,\mathrm{nm}, \ \ z=50\,\mathrm{nm}, \ \
W=16\,\mathrm{nm}\nonumber\\
&&\VPG-\VTG=4\,\mathrm{V}, \ \ \delta_A=0.
\end{eqnarray}
The difference between minimum and maximum potential in this case
is approximately 65~mV.
In the panel (b) of Fig.\ref{UxyId100} we present map of the electrostatic
potential for the  gate system with the same parameters,
but with some area disorder.
\begin{eqnarray}
\label{par2}
&&L=100\,\mathrm{nm}, \ \ D=50\,\mathrm{nm}, \ \ z=50\,\mathrm{nm}, \ \
W=16\,\mathrm{nm}\nonumber\\
&&\VPG-\VTG=4\,\mathrm{V}, \  \ \delta_A=2.5\,\mathrm{nm}
\end{eqnarray}
The rms area variation of the anti-dot, Eq.~(\ref{arms}),  is just 5\%.
Fig.\ref{UxyId100}(b) clearly demonstrates the disorder induced
puddles. The random variation of the potential is so strong that
the range of the potential variation in panel~(b) is almost
four times larger than that in panel~(a).
So, in this case the gate disorder completely kills the
miniband structure.

\begin{figure}[ht]
\includegraphics[width=0.4\textwidth,clip]{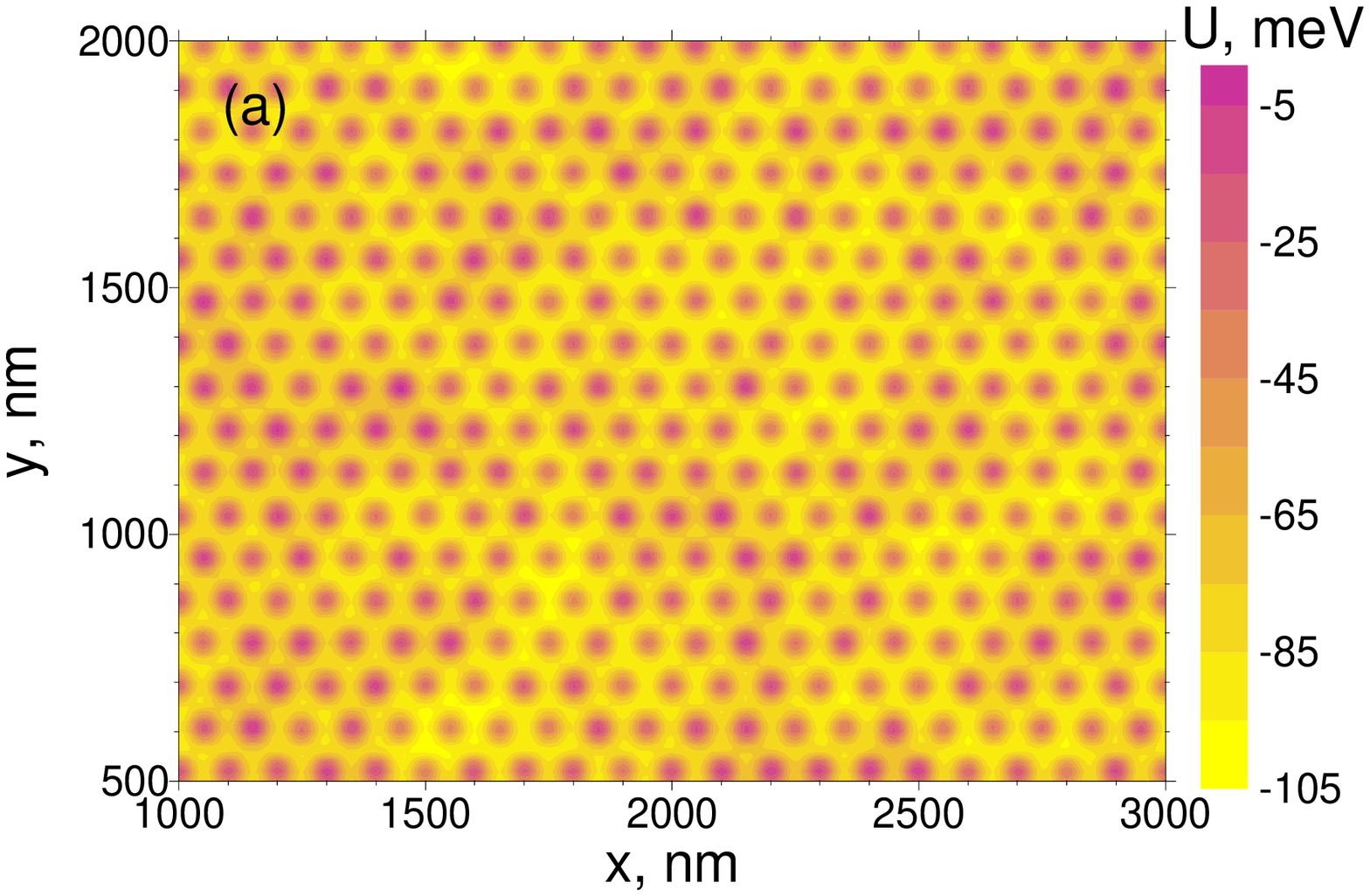}
\includegraphics[width=0.4\textwidth,clip]{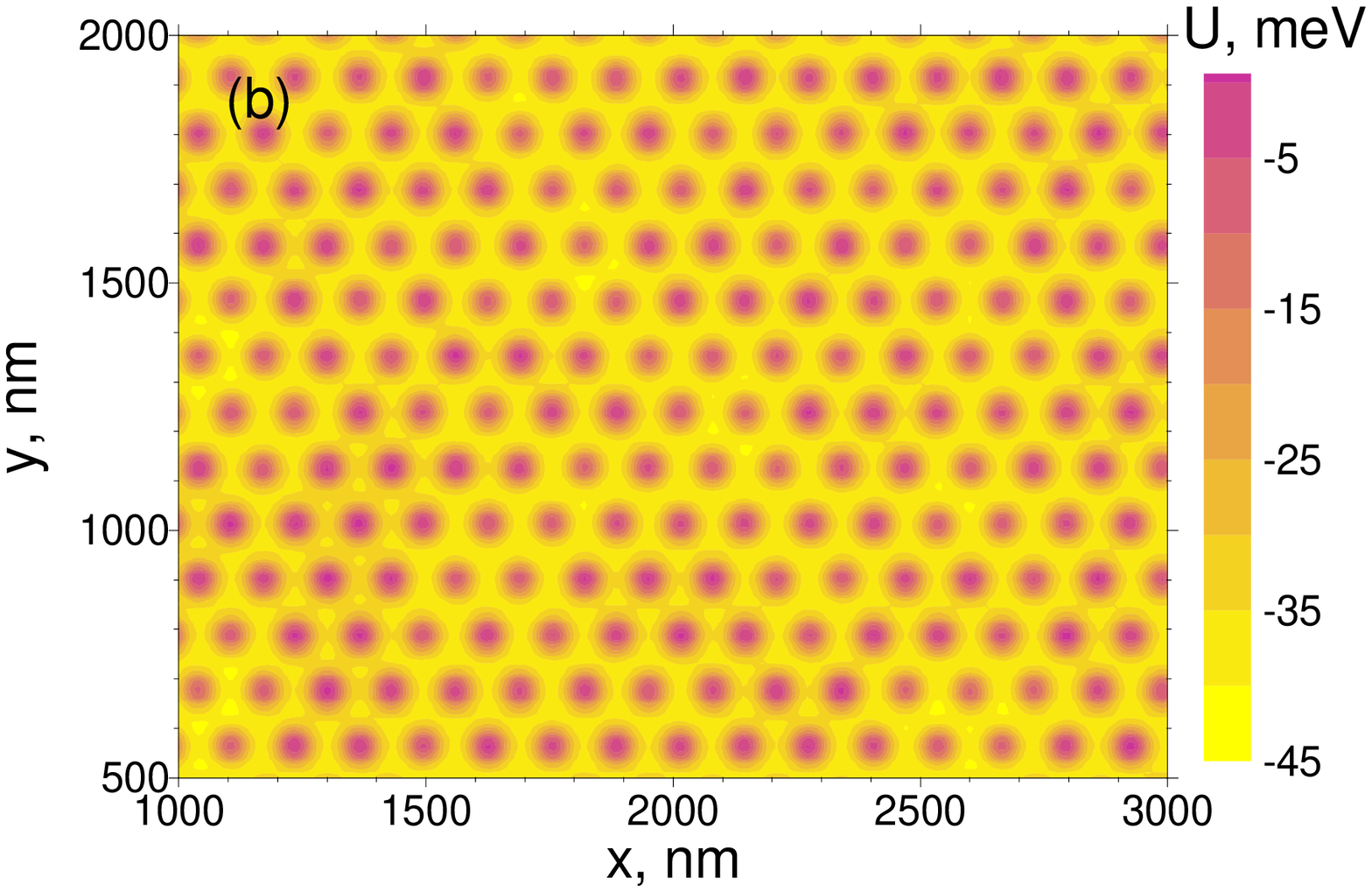}
\caption{Maps of the potential energy in the 2DEG plane
for  particular realizations of the gate disorder.
In panel~(a) the heterostructure parameters are
$L=100$\,{nm}, $D=50$\,{nm}, $z=37$\,{nm}, $W=16$\,{nm},
$\VPG-\VTG=1.6$\,V, $\delta_A=2.5$\,nm.
In panel~(b) the lattice spacing is somewhat increased
$L=130$\,{nm}, $D=60$\,{nm}, $z=37$\,{nm}, $W=16$\,{nm},
$\VPG-\VTG=0.44$\,V, $\delta_A=2.5$\,nm.
Both panels correspond to 5\% rms variation of the area of gate
system anti-dots.
}
\label{UxyDis}
\end{figure}

To reduce the relative effect of disorder we should decrease distance to
the gates $z$ and increase the lattice spacing $L$.
In panel~(a) of Fig.\ref{UxyDis} we present map of the electrostatic
potential for the  gate system with parameters similar to that
of Fig.\ref{UxyId100}(b). We just reduce $z$ and  accordingly reduce
$\VPG-\VTG$.
\begin{eqnarray}
\label{par3}
&&L=100\,\mathrm{nm}, \ \ D=50\,\mathrm{nm}, \ \ z=37\,\mathrm{nm}, \ \
W=16\,\mathrm{nm}\nonumber\\
&&\VPG-\VTG=1.6\,\mathrm{V}, \  \ \delta_A=2.5\,\mathrm{nm}
\end{eqnarray}
Puddling in Fig.\ref{UxyDis}(a)
is significantly smaller than that in Fig.\ref{UxyId100}(b),
but still it is too strong.
Finally in panel~(b) of Fig.\ref{UxyDis} we present map of the
electrostatic  potential for the  gate system with larger $L$
and accordingly reduced $\VPG-\VTG$.
\begin{eqnarray}
\label{par4}
&&L=130\,\mathrm{nm}, \ \ D=60\,\mathrm{nm}, \ \ z=37\,\mathrm{nm}, \ \
W=16\,\mathrm{nm}\nonumber\\
&&\VPG-\VTG=0.44\,\mathrm{V}, \  \ \delta_A=2.5\,\mathrm{nm}
\end{eqnarray}
Puddling in Fig.\ref{UxyDis}(b) is so small that it is not seen on
the potential map.

\begin{figure}[ht]
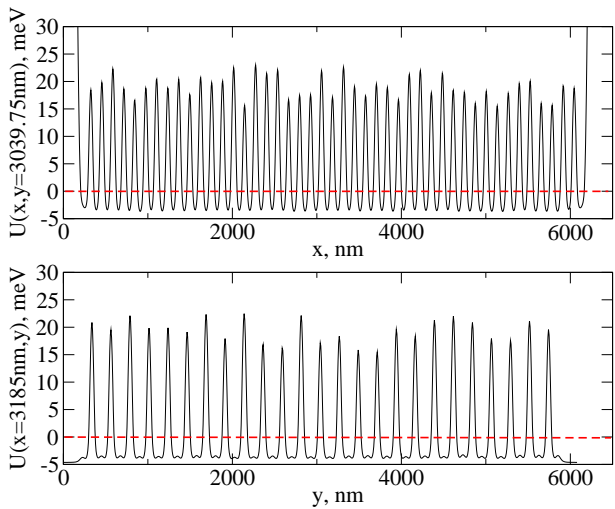

\includegraphics[width=0.45\textwidth,clip]{fig11a_uxd.eps}
\includegraphics[width=0.45\textwidth,clip]{fig11b_uyd.eps}
\caption{Self-consistent potential energy for disordered
$6\mum\times6\mum$ artificial graphene with parameters
presented in Eq.(\ref{par4}).
The potential energy is plotted
along the $x$- and the $y$-directions.
The chemical potential shown by the red dashed line is zero.
}
\label{uxyd}
\end{figure}
\begin{figure}[ht]
\includegraphics[width=0.23\textwidth,clip]{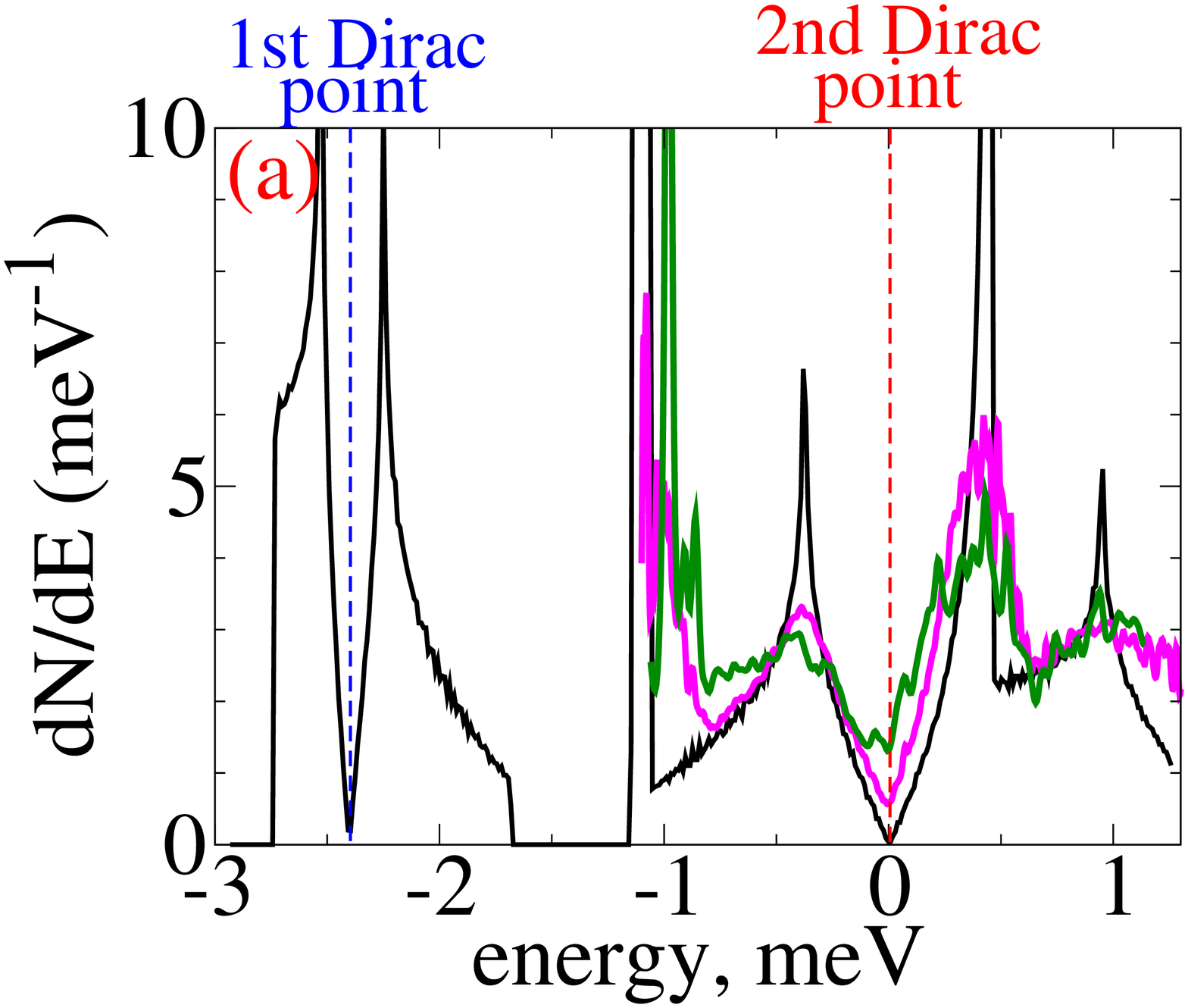}
\includegraphics[width=0.23\textwidth,clip]{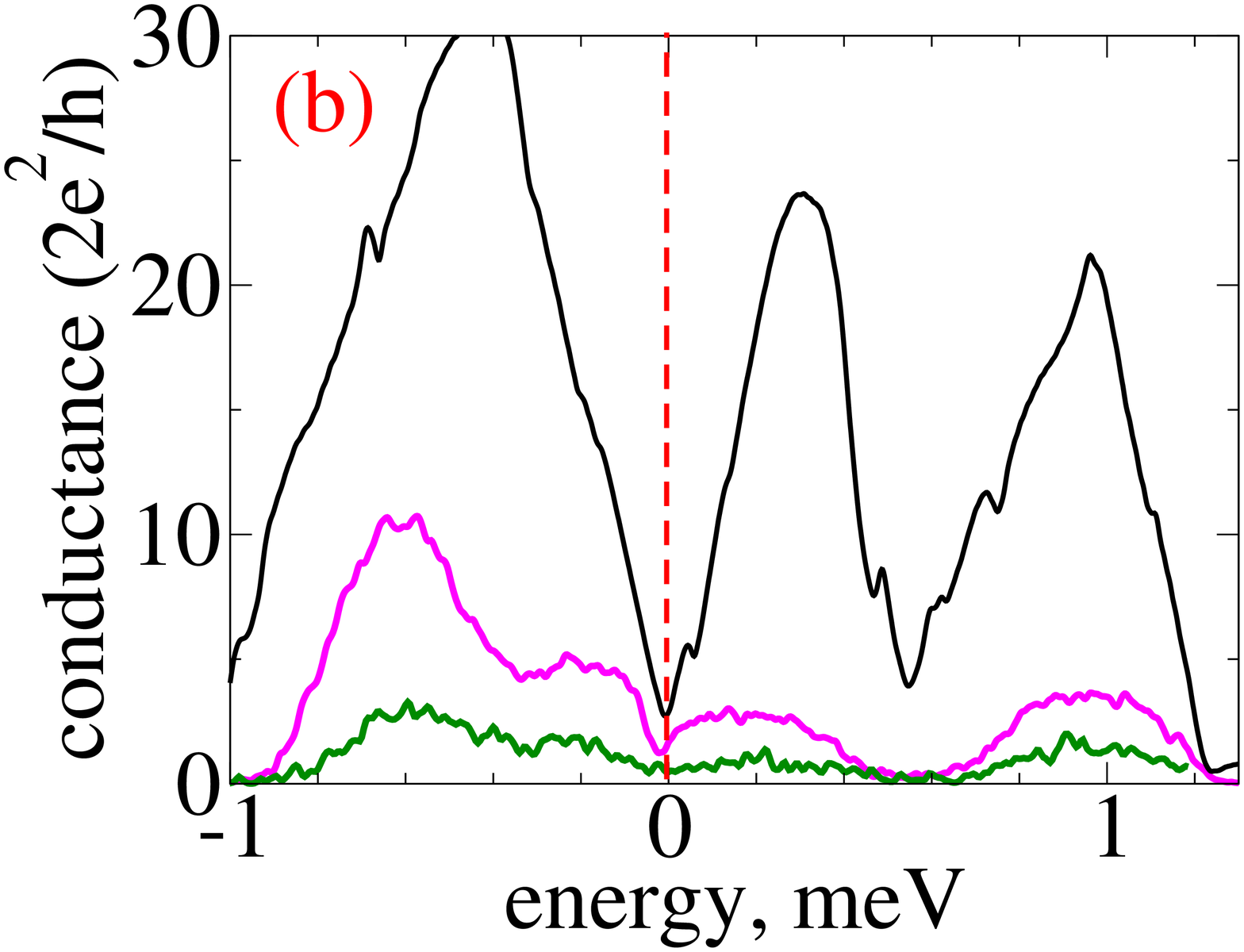}
\caption{(a) Density of states per  superlattice unit cell.
(b) Conductance of the artificial graphene
``sample'' of $6\mum\times6\mum$ size.
Parameters of the heterostructure are
$L=130$\,{nm}, $D=60$\,{nm}, $z=37$\,{nm}, $W=16$\,nm,
$\VPG-\VTG=0.44$\,V.
Black solid lines, identical to that in Fig.~\ref{dos}, show results
for the perfect sample, no disorder $\delta_A=0$.
Magenta solid lines show results for $\delta_A=2.5$\,nm which corresponds
to 5\% rms variation of the quantum anti-dot area.
Green solid lines show results for $\delta_A=5$\,nm which corresponds
to 10\% rms variation of the quantum anti-dot area.
}
\label{dosd}
\end{figure}

From Fig.\ref{UxyDis}(b) one concludes that
 the parameter set (\ref{par4}) is acceptable.
 The maps Fig.\ref{UxyId100} and Fig.\ref{UxyDis}
do not account for the 2DEG self screening.
The 2DEG self screening taken into account via Poisson-TF method
makes the potential fluctuation even smaller.
Plots of the final self-consistent potential for a particular realization
of the area disorder with parameter set (\ref{par4}) are presented in
Fig.\ref{uxyd}.
Note that leveling of the potential minimums in Fig.\ref{uxyd} is much better
than leveling of potential maximums. This is the effect of the charge density
redistribution over the connected charged manifold, the ``bright''
net in Fig.\ref{F5}. Such screening is more efficient in the anti-dot
geometry which we use. Leveling of the potential maximums is not as good
since there are no electrons in vicinity of the maximums.
The density of states and the conductance calculated  with this self-consistent
potential are  shown in Fig.\ref{dosd}(a),(b) by solid magenta lines.
The density of states clearly demonstrates minimum near the second Dirac
point, so the
set of parameters (\ref{par4}) is acceptable for the artificial graphene design.
On the other hand the dip in the conductance near the Dirac point is very weak.
This is not surprising, the size of the ``sample'' is sufficiently large,
and hence the electron dynamics becomes diffusive.
This is the reason why the entire conductance curve lies significantly
lower than that for the ideal sample  (black solid line in Fig.\ref{dosd}(b)).
Obviously, the diffusive regime itself does not invalidate the Dirac physics.

We have also increased the disorder using the same parameter set (\ref{par4})
with only one change, $\delta_A=2.5$\,nm $\to$ $\delta_A=5$\,nm.
The increased $\delta_A$  corresponds to 10\% rms variation of the anti-dot
area,  Eq.(\ref{arms}).
The density of states and the conductance calculated  in this situation
 are  shown in Fig.\ref{dosd}(a),(b) by solid green lines.
The density of states minimum near the second Dirac point has almost gone,
so this degree of disorder is  the border-line
for observation of the Dirac physics.
Interestingly, even at this degree of disorder the band structure
(allowed and forbidden energy bands) is still there and hence it can be observed in the density of states capacitance measurements \cite{Ponomarenko2010}.

\section{Conclusions\label{sec:conclusions}}
We have analysed effects of the Coulomb many-body screening and effects of
the residual disorder on the artificial graphene based
on two-dimensional electron gas.
Specifically we consider
AlGaAs/GaAs/AlGaAs heterostructure with two metallic gates.
The metallic gates make the many-body screening problem significantly
different from that in natural graphene.
We found that the design least susceptible to the disorder
corresponds to the weak coupling regime (opposite to tight binding)
which is realised via system of quantum anti-dots.
The most dangerous type of disorder is the area disorder which is
a random variation of areas of quantum anti-dots.
The area disorder results in formation of puddles.
Other types of disorder, the position disorder and the shape disorder,
are practically irrelevant.
The formation/importance  of puddles dramatically depends on parameters of the
nanopatterned heterostructure. For example a variation of the depth
of the heterostructure by 30\% (50\,nm $\to$ 37\,nm) in combination
with variation of the superlattice period by 30\% (100\,nm $\to$ 130\,nm)
results in suppression of the relative amplitude of puddles by almost two
orders of magnitude. Based on this analysis we formulate criteria for
the acceptable design of the nanopatterned heterostructure aimed at
creation of the artificial graphene.

\acknowledgements
We thank A. R. Hamilton and O. Klochan for interest to the work and for
numerous  discussions. We also thank D. Baksheev for software optimizations.
O.~A.~T. gratefully acknowledges the  School of Physics at the University of
New South Wales for warm hospitality during her visit.
O.~P.~S. gratefully acknowledges  Yukawa Institute for Theoretical Physics
for warm hospitality during work on this project.
This work was supported by Australian Research Council's Discovery
Project DP110102123. The work has been also supported by RAS
(Russian Academy of Sciences)  Presidium Program \#24.
The work has been also supported by SB RAS  Integration Project \#130.
Computations were performed on RAS Joint Supercomputing Center MVS-10P (www.jscc.ru).


\end{document}